\documentclass[%
 amsmath,amssymb,
 aps,
 prfluids,
]{revtex4-2}

\usepackage{tikz}
\usepackage{lineno,hyperref}
\usepackage{epsfig}
\usepackage{amssymb}
\usepackage{mathtools}
\usepackage{booktabs} 
\usepackage{enumitem}
\usepackage{graphicx}
\usepackage{subcaption}
\usepackage{pgfplots}
\modulolinenumbers[5]
\usepackage{dcolumn}
\usepackage{comment}
\usepackage{bm}
\usepackage[percent]{overpic}
\DeclareRobustCommand\dashed{\tikz[baseline=-0.6ex]\draw[thick,dashed] (0,0)--(0.54,0);}

\begin{document}


\title{The effect of $Re_\lambda$ and Rouse numbers on the settling of inertial particles in homogeneous isotropic turbulence }
\affiliation{
Univ Grenoble Alpes, CNRS, Grenoble-INP, LEGI, F-38000, Grenoble, France
}%
\author{Daniel Odens Mora}
 \altaffiliation[Also at ]{Department of Mechanical Engineering, University of Washington, Seattle, Washington 98195-2600, USA}
\author{Martin Obligado}
\email{Martin.Obligado@univ-grenoble-alpes.fr}
\affiliation{
University Grenoble Alpes, CNRS, Grenoble-INP, LEGI, F-38000, Grenoble, France}%

\author{Alberto Aliseda}%

\affiliation{%
 Department of Mechanical Engineering, University of Washington, Seattle, Washington 98195-2600, USA
}%

\author{Alain Cartellier}
\affiliation{
University Grenoble Alpes, CNRS, Grenoble-INP, LEGI, F-38000, Grenoble, France
}%

\date{\today}

\begin{abstract}

We present an experimental study on the settling velocity of dense sub-Kolmogorov particles in active-grid-generated turbulence in a wind tunnel. Using phase Doppler interferometry, we observe that the modifications of the settling velocity of inertial particles, under homogeneous isotropic turbulence and dilute conditions $\phi_v\leq O(10)^{-5}$, is controlled by the Taylor-based Reynolds number $Re_\lambda$ of the carrier flow.  On the contrary, we did not find a strong influence of the  ratio between the fluid and gravity accelerations (i.e., $\gamma\sim(\eta/\tau_\eta^2)/g$) on the particle settling behavior. Remarkably, our results suggest that the  hindering of the settling velocity (i.e. the measured particle settling velocity is smaller than its respective one in still fluid conditions) experienced by the particles increases with  the value of $Re_\lambda$, reversing settling enhancement found under intermediate $Re_\lambda$ conditions. This observation applies to all particle sizes investigated, and it is consistent with previous experimental data in the literature. At the highest $Re_\lambda$ studied,  $Re_\lambda>600$, the particle enhancement regime  ceases to exist. Our data also show that for moderate Rouse numbers, the difference between the measured particle settling velocity and its velocity in still fluid conditions scales linearly with Rouse, when this difference is normalized by the carrier phase rms fluctuations, i.e., $(V_p-V_T)/u\sim -Ro$. 
\end{abstract}

\maketitle


\section{Introduction}

Turbulent particle-laden flows have a widespread presence in industrial and natural processes, e.g., coatings, spray combustion, pollen dispersion, planetesimal  growth, and clouds formation \cite{aliseda2006, vaillancourt2000review,sumbekova2017preferential}. Among the several consequences of particle-turbulence interactions, preferential concentration and particle settling velocity modification have received considerable attention in the last decades \cite{balachandar2010turbulent,aliseda2011, elghobashi2019direct}. Preferential concentration describes the tendency of particles to accumulate in space, forming clusters and voids. In contrast, particle settling modification accounts for the enhanced (resp. hindering) particles settling velocity in the direction of a body force acting on them, for instance, gravity. 

Several theoretical approaches have suggested mechanisms that relate the topology of the turbulent flow to the observed phenomena. Classical contributions have suggested that \textit{sub-Kolmogorov} particles, which have a characteristic scale smaller than the Kolmogorov scale $\eta$,  tend to concentrate in regions of high strain and low vorticity \cite{maxey1987gravitational,balachandar2010turbulent}. However, recent research has proposed that this classical picture does not take into account the multiscale nature of turbulence. Under this framework, some studies have proposed that particles accumulate at the different (coarse-grained) scales of high strain and low vorticity \cite{bragg2015mechanisms}. Alternatively, others have shown evidence that particles mimic the clustering of the carrier phase zero acceleration points \cite{coleman2009unified,Obligado2014}, which exhibits a self-similar behavior \cite{goto2006self}.

Previous studies have suggested that modification of particles settling velocity may be due to centrifugal effects in preferential sweeping: inertial particles are expelled of eddies but fast-track into downward eddies, thereby enhancing their settling speed \cite{maxey1987gravitational,ghosh2005turbulence,wang1993settling}. The opposite effect has also been observed: particles settling velocity is reduced instead of being enhanced \cite{nielsen1993turbulence}. Some research has conjectured that this phenomenon occurs when particles  prefentially sample the upward regions of the flow
~\cite{nielsen1984motion,nielsen1993turbulence,kawanisi2008turbulent}. Recent studies have also attempted to incorporate ideas from the multiscale nature of turbulence to understand the observed particle settling behavior. Some works argue  that the centrifugal effect (and enhanced settling) depends on the relationship between the particle inertia, and all of the carrier phase length scales, i.e., particles of different inertia are affected by different length scales of the turbulent flow \cite{tom2019multiscale}. 

Considering the complex interactions between the turbulent carrier phase, and the discrete particle phase, most studies treat preferential concentration and particle settling independently. Recent research  \cite{Sumbekova2019,sumbekova2016clustering,monchaux2017settling,baker2017coherent,huck2018role,petersen2019experimental},however, has aimed at relating both phenomena. For instance, some studies have reported that the enhanced particle settling is due to the increased local concentration \cite{aliseda2002effect,huck2018role}. In other words, particles in high density regions settle (on average) faster with respect to particles in low density regions \cite{aliseda2002effect,bosse2006small,coleman2009unified,good2014settling,bec2014gravity,ireland2016effect,dhariwal2018small}.

Numerical and experimental studies do exhibit similar trends on the behavior of preferential concentration and settling velocity with global flow parameters, such as the Taylor-based Reynolds number $Re_\lambda=u\lambda/\nu$ and the Stokes number $St=\tau_p/\tau_\eta$. $u$ stands for the RMS value of the streamwise fluctuating velocity $u'$, $\lambda$ corresponds to the Taylor microscale and $\nu$ to the kinematic viscosity. $\tau_p$ and $\tau_\eta$ stand for the particle relaxation and the Kolmogorov timescales, respectively. Nevertheless, quantitative consensus has yet to be reached \cite{bosse2006small,monchaux2017settling,wittemeier2018explanation,ireland2016effect,tom2019multiscale,petersen2019experimental}. Moreover, the origin of these discrepancies could be multi-fold \cite{rosa2016settling,bosse2006small,good2014settling,monchaux2017settling}. The numerical and experimental study of Good et al. \cite{good2014settling}, for example, at similar values of  $Re_\lambda $, and $\phi_v$ have shown that particle settling \textit{hindering} effects cannot be captured in DNS simulations that only consider linear drag. Conversely, DNS simulations of Rosa et al. \cite{rosa2016settling}  report no variation in the particle settling velocity with the drag model, i.e.,  their results were insensitive to the choice of the drag law used (e.g., linear, non-linear).

Another source of discrepancy may stem from the mechanical coupling between particle phase and the turbulent carrier phase interaction  ignored in most DNS studies. 
 The need to include these inter-phase mechanical coupling effects was recognized early by Aliseda et al. \cite{aliseda2002effect}. They suggested modifying the carrier phase pressure field to account for the flow regions with high particle density.  Most DNS studies ignore this coupling and assume that the particles do not affect the carrier phase,  a regime known as `one-way' coupling. However, Bosse et al. \cite{bosse2006small},  and Monchaux et al. \cite{monchaux2017settling} observed a larger particle settling velocity when there is mechanical coupling between the phases, a  regime known as `two-way' coupling. Their simulations, however, were run at rather small Reynolds numbers ($Re_\lambda\approx40$). Rosa et al. \cite{rosa2020effects} have recently arrived at similar conclusions at higher Reynolds numbers ($Re_\lambda\approx100$). These findings hint that including two-way coupling interactions may be necessary to describe accurately the physics underlying this phenomenon.

In this work, we report experimental measurements of a polydisperse population of inertial particles settling under homogeneous isotropic turbulence downstream of an active grid. For $Re_\lambda \in [230-650]$, our results suggest that the Taylor-Reynolds number ($Re_\lambda$) is the leading contributor to the particles' behavior, influencing all the measurable regimes. For instance, the degree of hindering (i.e. measured particle settling velocity smaller than its respective value in still fluid) increases with the value of $Re_\lambda$. Moreover, the transition point between hindering and enhancement (particles falling faster than in a quiescent fluid) regimes shifts to smaller Rouse numbers at increasing values of $Re_\lambda$.


\section{Methods}
\subsection{Experimental Setup}
\label{sc:exps}

The experiments were performed in a close-circuit wind tunnel `\textit{Lespinard}' in the \textit{Laboratoire des \'{E}coulements G\'{e}ophysiques et Industriels} (LEGI) at  Universit\'{e} Grenoble Alpes. This facility has been regularly employed to study particle clustering under \textit {Homogeneous Isotropic Turbulence} (HIT) conditions \cite{Monchaux2010,obligado2015experimental,sumbekova2017preferential,mora2020estimating}. A sketch of our experimental setup is depicted in appendix \ref{sca:exps} (see figure \ref{fig:WT_sk}). In our experiments, the turbulent flow was generated utilizing an active grid \cite{mydlarski2017turbulent} in triple random mode. We measured the turbulent unladen velocity through hot-wire anemometry. We computed the turbulent parameters using standard methods and assumptions (e.g., Taylor hypothesis). The most relevant parameters are summarized in table \ref{tab:param}. For detailed explanations on the turbulence characterization, see \cite{MoraPRF2019}. Figure \ref{fig:spk} shows the energy spectra at the measuring station (see label `\textbf{M1}'  in appendix \ref{sca:exps} figure \ref{fig:WT_sk}). 

Droplets were injected right behind the active grid using a rack of injectors  (see figure \ref{fig:WT_sk} in appendix \ref{sca:exps}). Injected droplets, with diameters $D_p$ between 20 and 300 microns, are considered spherical, as their Weber number is below unity (see section 6.3 in \cite{Sumbekova2016a} ).  We measured the particles diameter, and the horizontal and vertical components of the velocity at 3m downstream of the active grid by means of phase Doppler interferometry (PDI) \cite{bachalo1984phase}. For each experimental condition, we collected data from $5\times10^{5}$ particles. The vertical (resp. horizontal) velocity component had a resolution of 0.010 m/s (resp. 0.04 m/s) for all experimental conditions.

The choice of the measurement position (3m downstream of the injection) is based on previous studies at the same facility. These studies recover that, at 3m downstream of the injection, the particle statistics are almost Gaussian. Hence, our measurements are able to gauge the effects of the background turbulence on the particle behavior. For more details on the  experiment, see appendix \ref{sca:exps}, and Sumbekova \cite{Sumbekova2016a}. On the other hand, a recent study in our facility reveals that carrier phase turbulence may change due to the particles presence \cite{MoraPRL2019}. To control the influence of such turbulence modulation due to the particles' presence, we ran the experiments with the smallest liquid fractions attainable in our facility\cite{Sumbekova2016a} (i.e. $\phi_v =[10^{-6}, 10^{-5}]$). We expect that, at these liquid fractions, the turbulence modulation is minimal \cite{elghobashi1994predicting}. 

All these previous considerations led to the exploration of a parameter space aiming at small concentrations and large Reynolds numbers, exploiting the limits of the facility (see figure \ref{fig:param}). 
\begin{figure}
	\begin{center}

		\begin{subfigure}[t]{0.48\textwidth}
		\begin{overpic}[scale=0.5]{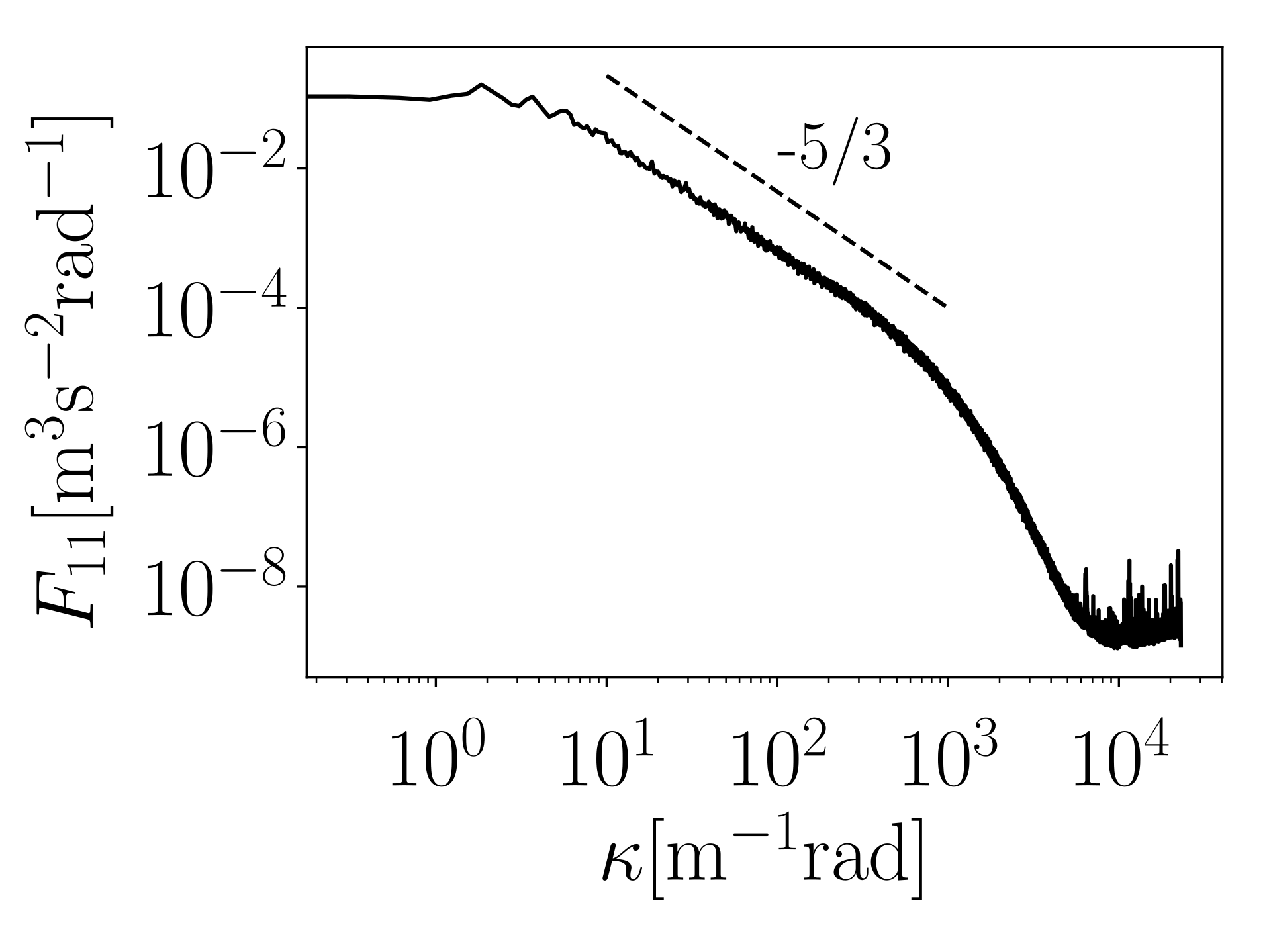}
			             \put (85,60) {\huge a)}
            \end{overpic}
			\caption{\label{fig:spk}}
		\end{subfigure}
		~
		\begin{subfigure}[t]{0.48\textwidth}
			\begin{overpic}[scale=0.5]{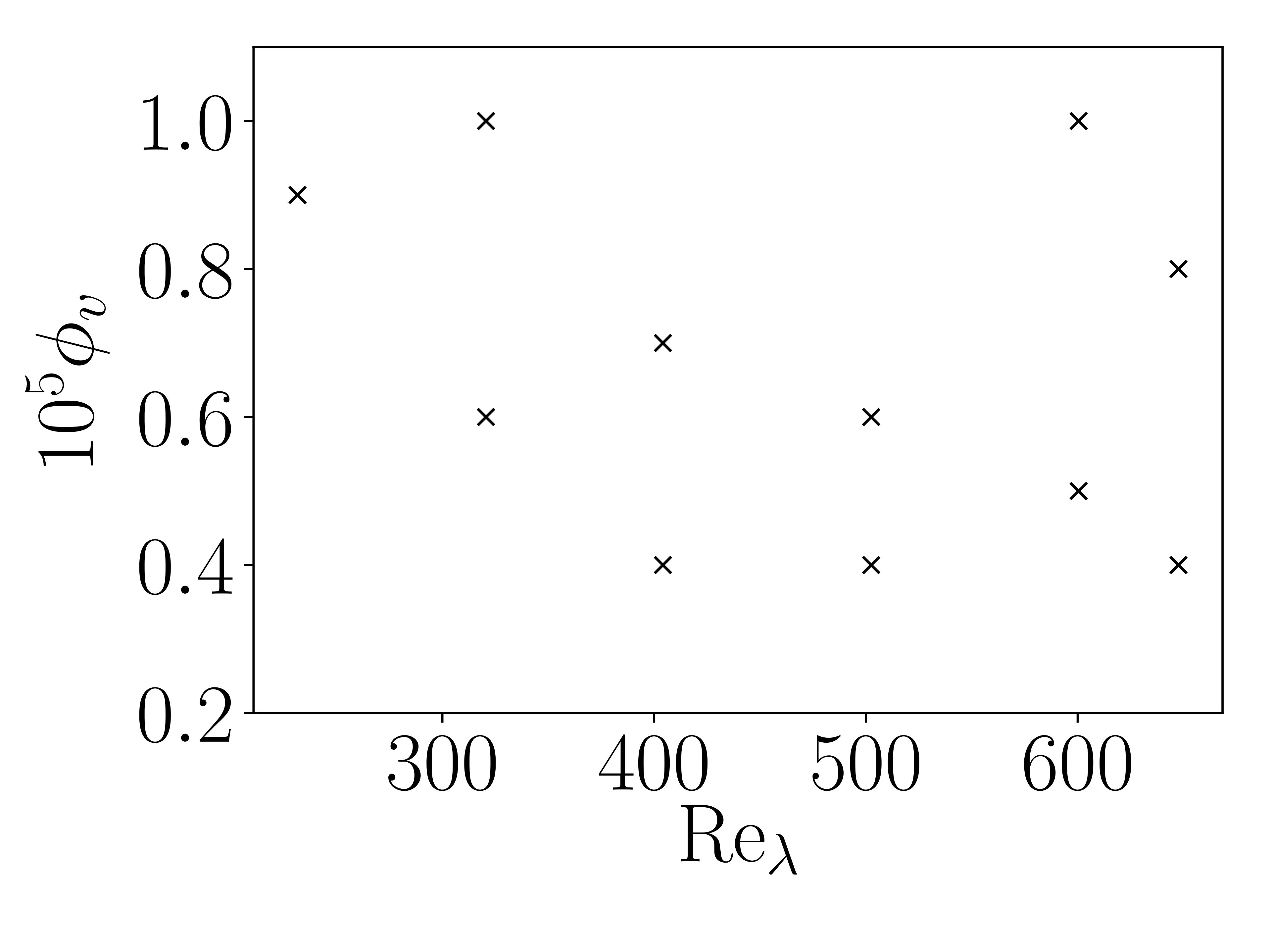}
    \put (25,25) {\huge b)}
            \end{overpic}			
            \caption{\label{fig:param}}
		\end{subfigure}
	\end{center}
	\caption{a) Energy spectrum example from hot-wire records at measuring station \textbf{M1} (see figure \ref{fig:WT_sk}). b) Parameter space for the experiments conducted. The global liquid fraction was estimated as $\phi_v\approx Q_W/Q_A$, where $Q_W$, and $Q_A$ are the volumetric flux of water, and of air, respectively. }	
\end{figure}

\begin{table}	
	\begin{center}
		\begin{tabular}{llcccccc}
			$Re_\lambda$&$U_\infty$& $u/U_\infty$ & $L$ &$\varepsilon$ & $\lambda $ & $\eta$& $\gamma$ \\
			&[ms$^{-1}$]&&[cm]&[m$^{2}$s$^{-3}$]&[cm]&[$\mu$m]& \\ \hline
			232&      2    &0.1273 & 5.70&0.0777& 1.36& 457  &    0.24\\
			321&      3    &0.1343 &  7.21 &0.2577&1.19  &338 & 0.59  \\     
			404&      4    &0.1405  &8.45&0.6058 &1.08&273& 1.12     \\
			503&      5    &0.1476 &9.80&1.1667&1.02&   231&1.84    \\  
			601&      6    &0.1541  &11.10 &2.1116&0.98  &  200& 2.87  \\     
			648&       7   &0.1578 &11.58 &3.3862&0.90 &   178&4.09   \\\hline

		\end{tabular}
	\end{center}
	
	\caption{Parameters of the unladen flow, measured by means of hot wire anemometry, at the measuring station 3 m downstream of the grid. The parameters are defined as: $u=\langle u^\prime\rangle^{1/2}$, the turbulence energy dissipation rate $\varepsilon=15\nu u^2/\lambda^2 $, $\eta=\big(\nu^3/\varepsilon\big)^{1/4}$, and $L$  is the integral length scale computed following \cite{puga2017normalized}. The kinematic viscosity of the air is taken as $\nu=1.5\times 10^{-5}$. Finally, $\gamma=\varepsilon^{3/4}/(g\nu^{1/4})$ is the acceleration ratio.  \label{tab:param}} 
\end{table}

\subsection{Velocity measurements and angle correction}

There will always be a small deviation angle between the PDI axes, and the wind tunnel frame of reference (see figure \ref{fig:fang} ) impacting the vertical velocity measurements.  Considering that the particles' horizontal velocity is at least an order of magnitude larger than the vertical one, the horizontal component's projection onto the vertical component in the PDI frame of reference will cause an error in the vertical velocity measurements due to optical misalignment. We address this problem by subtracting the projected mean droplet horizontal ($\langle U_p\rangle$) velocity from the vertical velocity in the PDI frame of reference ($ V_p$). Thus, we define the angle-corrected velocity as:

\begin{equation}
V_{j}^c=V_{j}-\langle U_p \rangle \mathrm{sin}(\beta) = V_{j}-V_\beta ,\quad V_\beta = \langle U_p \rangle \mathrm{sin}(\beta)
\label{eq:disa}
\end{equation}

To estimate $V_\beta$, we used a different configuration in the wind tunnel. We used a single particle injector, positioned at the grid plane and set the grid completely open, thus minimizing turbulence. We circulated air at 3.5 $ms^{-1}$ and injected olive oil droplets, with a very narrow distribution of sizes, centered around a mean diameter $\approx 8 \mu m$ (measured with the PDI). Particles were therefore convected downstream. The settling velocity of droplets can be estimated via the Schiller and Nauman \cite{clift1978bubbles} drag coefficient semiempirical formula.

We then measured these droplets' velocities at the PDI measuring volume (see \ref{fig:WT_sk}). The velocity statistics collected for 2000 droplets in the PDI frame of reference (see figure \ref{fig:fang}) were $\langle U_p\rangle = (-3.52\pm0.02)m/s$, $\sigma_{U_p}=(0.11\pm0.02)m/s$, and $\langle V_p\rangle=(-0.09\pm 0.005)m/s$, $\sigma_{V_p}=(0.11\pm0.005)m/s$. The latter values, the Schiller and Nauman formula and our resolution yielded $\beta=-1.5^\circ\pm0.3^\circ$, a correction angle we used for all experimental realizations. The angle uncertainty comes from the PDI velocity resolution.

The angle correction for all particles is justified under our turbulent conditions because this correction is smaller than the standard deviation of the carrier phase velocity, i.e.,
 $V_\beta/ u=\mathrm{sin}(\beta)\times \langle U_p \rangle/u\approx\mathrm{sin}(1.5^\circ)\times O(100)<1$ (see table \ref{tab:param}, and figure\ref{fig:velpdf} ).

{\color{red} \begin{figure}[h!]
	\includegraphics[scale=0.45]{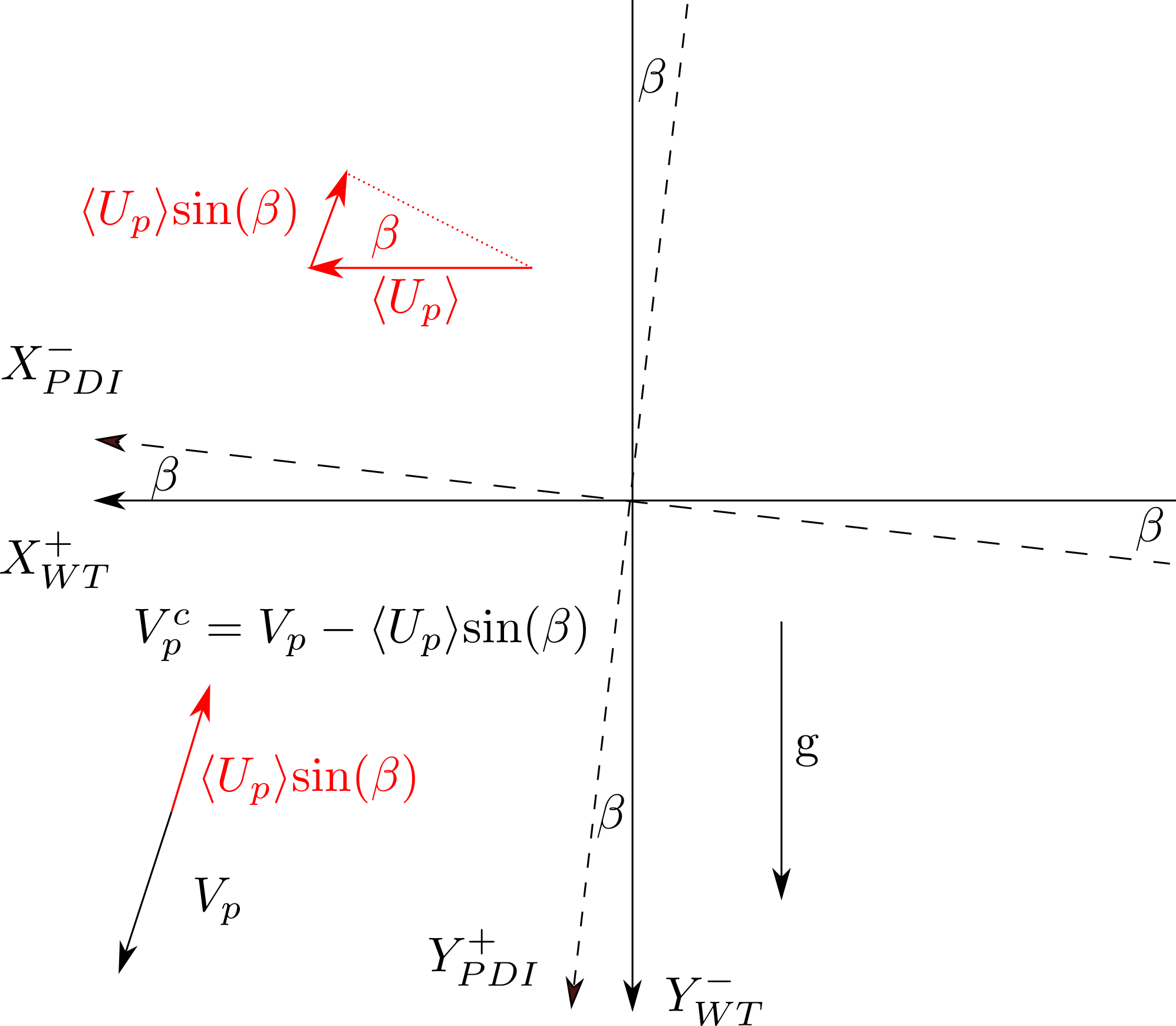}

	\caption {Frame of reference for the wind tunnel, and the PDI device. \label{fig:fang}}
	
\end{figure}
}

\section{Measurements}
\subsection{Raw settling velocity}

We will consider the particles' vertical velocity to be positive towards gravity consistent with the PDI frame of reference (figure \ref{fig:fang}). We binned our datasets by the droplet diameters. These bins had a size of 5$\mu m $ (an operation represented by $\langle \rangle \vert_D$) and their centers spanned $D_p\in[7.5-148.5] \mu$m. This latter consideration is due to the injector droplet size distribution and has some consequences: smaller droplets are less common (see figure \ref{fig:spray_pdf}), and therefore, our first bins have a larger variation. We, nevertheless, collected enough samples to have meaningful statistics.

\begin{figure}[h]
	\centering

		\includegraphics[scale=.25]{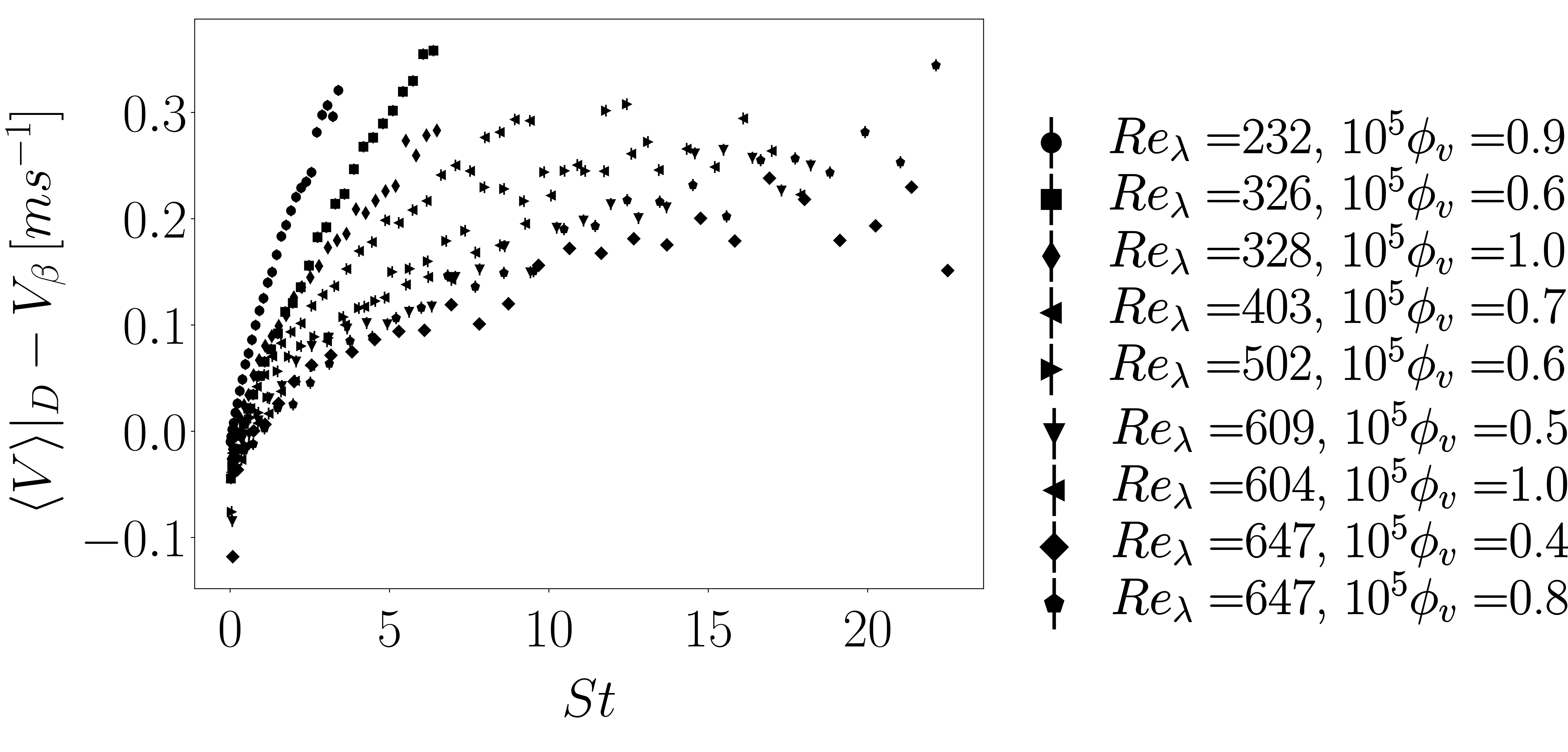}

	\caption{ Particle vertical velocity measurements binned by diameter size against the binned Stokes number. Error bars have a size $\pm5\times10^{-3}ms^{-1}$ (half of the PDI resolution). \label{fig-raw-st}}

\end{figure}

Our raw velocity measurements show that for a fixed experimental condition, as expected, particles with larger Stokes (larger diameters in our case) fall --on average-- faster (see figure \ref{fig-raw-st}). However, there are two sources of uncertainty in our results for the smallest droplets: first, the accuracy of the optical alignment and, second, the vertical resolution used (0.010 m/s). In the latter, the resolution results from a trade-off between an adequate acquisition rate and the statistics needed. 

Interestingly, for all particle sizes, the particles' velocities decrease with increasing $Re_\lambda$ (i.e., slower settling in our convention). The polydispersity of our droplet injection and our active grid turbulence characteristics (e.g., higher values of $\varepsilon$, see table \ref{tab:param}) allows us to explore a wide range of particle Stokes numbers for the different experimental conditions (see figure \ref{fig-raw-st}). 

\subsection{Settling parameters, and non-dimensional numbers}
\label{sc:nondim}

The carrier phase is usually characterized by the Taylor Reynolds number $Re_\lambda=u\lambda/\nu$. Conversely, there is still an open debate (see \cite{Sumbekova2016a} and references therein) about which parameters are adequate to describe the dynamics of the discrete phase. A complete discussion on how to derive these parameters from dimensional analysis (or from first principles) is beyond the scope of this study (see section 6.5 in \cite{Sumbekova2016a} ). Thus, we briefly summarize the most common non-dimensional numbers proposed to analyze the particles settling velocity.

Classical numerical and experimental studies \cite{aliseda2002effect,wang1993settling} plot the particles settling velocity against the Stokes number $St=\tau_p/\tau_\eta$ (see figure \ref{fig-raw-st}); changes in the turbulence dissipation lead to changes in the Stokes number (figure \ref{fig-raw-st}).

Other non-dimensional parameters of interest involve the ratio between the particle terminal speed ($V_T$) and the background turbulence RMS fluctuation, known as the Rouse number, $Ro=V_T/u$ \cite{Sumbekova2016a,monchaux2017settling} (some authors also refer to this non-dimensional number as the settling parameter $Sv$ \cite{good2014settling,petersen2019experimental}).  Algebraic manipulations allow combining Rouse and Stokes numbers as follows:

\begin{equation}
St= \frac{\tau_p}{\tau_\eta}\rightarrow St= \frac{\tau_p  }{\tau_\eta }\frac{u}{u}\frac{g}{g}\rightarrow St=Ro \frac{u}{\tau_\eta  g},
\label{eq:St1}
\end{equation}

\noindent where the particle relaxation time includes the non-linear drag from Schiller and Nauman \cite{clift1978bubbles},
\begin{equation}
\tau_p=\frac{\rho_pD_p^2}{18\mu_f(1+0.15Re_p^{0.687})}, \quad V_T=\tau_p g.
\label{eq:chN}
\end{equation}

\noindent In addition to $St$ and $Ro$, some research suggest that the ratio between the turbulent acceleration ($\eta/\tau_\eta^2$) and gravity may play a role on the results. Some authors refer to this ratio as $\gamma= \eta/(g\tau_\eta^2)$ \cite{huck2018role,nielsen1993turbulence,good2014settling}, while others refer to it as a \textit{Froude} number \cite{bec2014gravity,tom2019multiscale} (Fr). In this work, we will follow the former notation. From equations (\ref{eq:St1} - \ref{eq:Fr}), and taking into account that $\lambda=\sqrt{15}\tau_\eta u$ (small scale isotropy), and that $u_\eta=\eta/\tau_\eta$, one gets;

\noindent 
\begin{minipage}{.3\linewidth} 
	\begin{equation}
	\gamma= \frac{\varepsilon^{3/4}}{g\nu^{1/4}}=\frac{\eta}{\tau_\eta^2g},
	\label{eq:Fr}
	\end{equation}
\end{minipage}\hfill
\begin{minipage}{.3\linewidth} 
	\begin{equation}
	St=\gamma\frac{Ro Re_\lambda^{1/2}}{15^{1/4}},
	\label{eq:St2} 
	\end{equation}
\end{minipage}\hfill
\begin{minipage}{.3\linewidth}
	\begin{equation}
	Ro= 15^{1/4}\frac{St}{\gamma Re_\lambda^{1/2}}.
	\label{eq:Ro} 
	\end{equation}
\end{minipage}
\vspace{0.5cm}

\noindent Moreover, combinations of these parameters such as $RoSt$ (involving the Rouse and Stokes numbers) have received interest recently, as they appear to give a better collapse of the data \cite{ghosh2005turbulence,good2014settling,petersen2019experimental}. For the $RoSt$, one gets from  equations \ref{eq:St2} and \ref{eq:Ro} that:

\begin{equation}
RoSt=\frac{V_T}{u}\frac{\tau_p}{\tau_\eta} \sim V_T  \frac{\tau_p}{\lambda},
\label{eq:Rost} 
\end{equation}
which seems to take into account the influence of the background turbulence on the particle settling velocity; the ratio between the particle stopping distance to the Taylor microscale $\lambda$, which scales with the average distance between velocity stagnation points \cite{liepmann1953counting,sreenivasan1983zero,mazellier2008turbulence,vassilicos2015dissipation,MoraPRF2019}. 

In our experiments, we cannot change the magnitude of the acceleration of  gravity ($g$) or the magnitude of the air kinematic viscosity ($\nu$). As a result, we cannot easily disentangle or individually vary, Ro, St, and $\gamma$. Therefore, we can only increase the turbulence dissipation $\varepsilon$ by increasing the inlet velocity $U_\infty$. These constraints yield similar functional behaviors for $\gamma$, and $Re_\lambda$. Thus, to overcome these restrictions, we complement and compare our results with other experimental datasets taken from different experimental studies.

\subsection{Normalized settling velocity}
\label{sc:nvm}
To quantify the degree of settling enhancement, the velocity difference  between the particle settling velocity and its terminal speed is computed, i.e., $\Delta V =\langle V\rangle\vert_D-V_T-V_\beta$, where $V_\beta$ includes the misalignment effects. $\Delta V$ is usually normalized by the carrier phase fluctuations $u$, or by the particle terminal speed $V_T$ \cite{wang1993settling,balachandar2010turbulent,aliseda2002effect,Sumbekova2019,petersen2019experimental,bec2014gravity,rosa2016settling}. 

\begin{figure}
	\centering
		\includegraphics[scale=0.18]{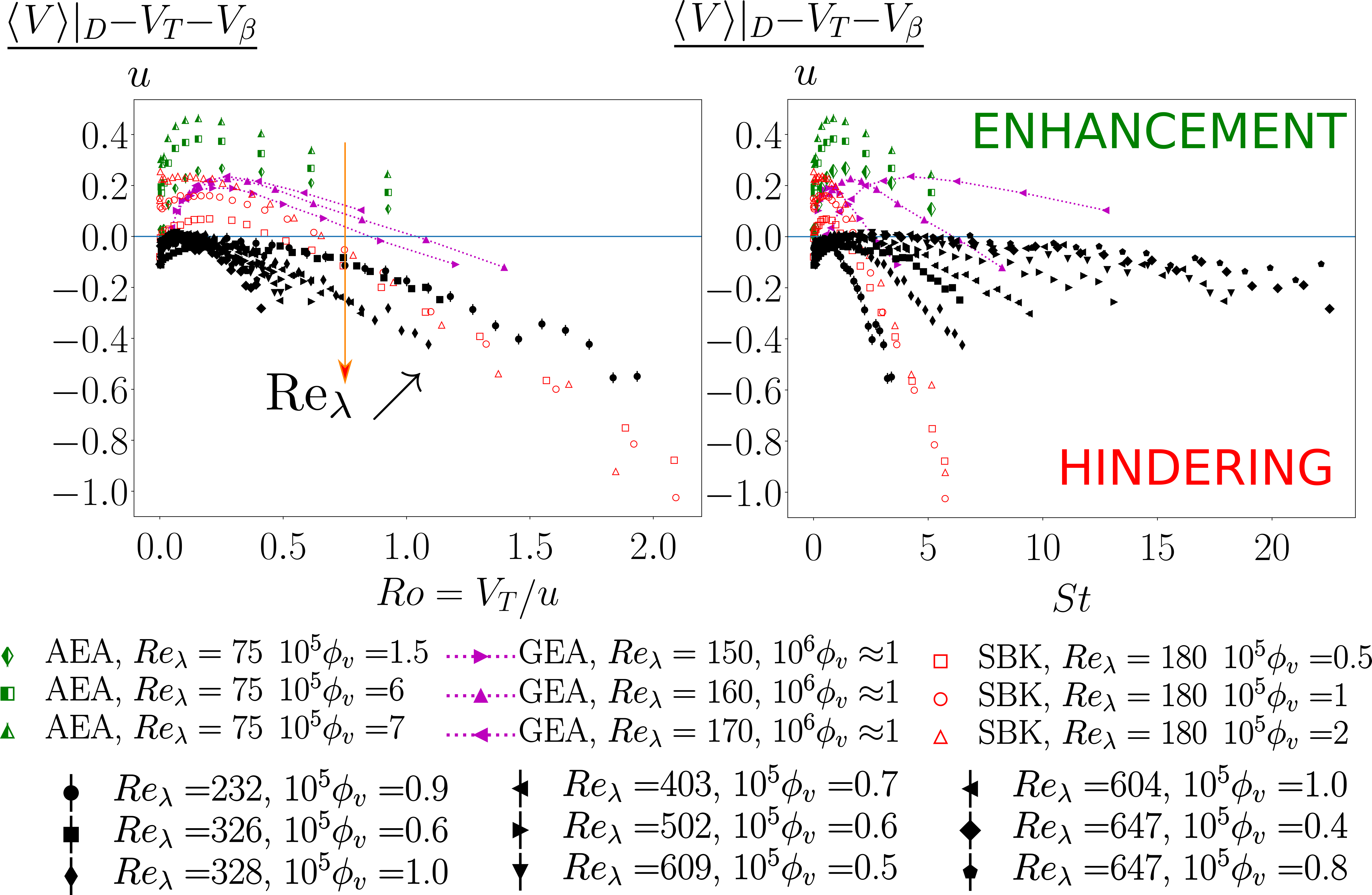}

	\caption{Particle velocity over the carrier phase fluctuations against Rouse (left) and Stokes numbers (right) . In the figures legend GEA refers to the data of Good et al.\cite{good2014settling}. AEA refers to the data of Aliseda et al. \cite{aliseda2002effect}, and SBK refers to the data of Sumbekova \cite{Sumbekova2019}. Error bars denote the resolution uncertainty. \label{fig-dv-ro}}
	
\end{figure}

Interestingly, previous experiments \cite{Sumbekova2019,good2014settling}, as well as ours, reveal that the particle velocity is hindered (slowed down with respect to the still fluid terminal velocity) as the $Re_\lambda$ increases above a certain threshold (see figure \ref{fig-dv-ro}). Other experiments, e.g., Akutina et al. \cite{akutina2020experimental} have also reported hindering for particles falling inside a turbulent column. Although, particles with small Rouse and Stokes numbers have settling velocities (magnitudes) that depend strongly on the liquid fraction $\phi_v$ and $Re_\lambda$, after the peak of maximum settling enhancement we observe that for $Ro>O(0.1)$, the normalized particle settling ($\Delta V/u$) seems to have a quasi-linear behavior (see figure \ref{fig-dv-ro}). To the authors best knowledge, this regime is not predicted by available analytical models. When $\Delta V/u$ is plotted against the Stokes number, we also observe that hindering is present at large $St$ numbers. The latter observations imply each other, due to the relationship between Rouse and Stokes numbers (c.f. equation \ref{eq:St2}). 

Our data exhibits hindering effects at very small $St$, and Rouse numbers, in agreement with findings in other experimental facilities, e.g., experiments in grid tanks \cite{jacobs2016flow}  and in a turbulence  box \cite{petersen2019experimental}. However, we must note that these conclusions require further research given the difficulty of recovering the `tracer' behavior in similar experimental measurements, i.e., a particle that almost perfectly follows a fluid parcel. To recover this behavior using laser interferometry (e.g. PDI) and imaging (e.g. PIV, PTV), it is required that the optical alignment is very accurate so that the absolute zero is adequately set. Besides proper alignment, we also need two extra elements: very dilute conditions $\phi_v \to0$, and, in our specific case, very small particles $St \to 0$. Thus, it is not surprising that most experiments have reported values of $\Delta V \neq 0$ for $St\to 0$ \cite{good2014settling,good2012intermittency,Sumbekova2019}.

Moreover, our measurement resolution could also have an impact on the measurements taken in the low $St$ regime. These resolution limitations can be clearly observed when the velocity $\Delta V$ is normalized against $V_T$ (in the appendix \ref{sc:scvt} see
figures \ref{fig-dvst-ro}, and \ref{fig-dvst-st} and the large error bars for small Rouse). We note that these conclusions could be biased by a condition that may exist due to the spatial domain where the experiments take place (confinement effects): weak recirculation currents that perturb the settling dynamics of the particles. These perturbations could be of the order of the settling velocity for small inertial particles biasing the results measured. These biases imply that the tracer behavior may not be recovered $\Delta V/V_T\neq0$ for $St\to 0$, and therefore, measuring the true values of  $\Delta V/V_T$ for $Ro\ll1$ or $St\ll1$ is not straightforward (see
figures \ref{fig-dvst-ro} and \ref{fig-dvst-st} in Appendix \ref{sc:scvt}). 

For instance,  Good et al. \cite{good2012intermittency} reports $\Delta V/V_T\to O(100)$ for $Ro\ll1$ in wind tunnel experiments. In a following publication, Good et al. \cite{good2014settling} suggest their previous experimental observation (i.e. $\Delta V/V_T\geq O(10)$ for $Ro\ll1$) was due to a weak mean flow. Likewise, Akutina et al. \cite{akutina2020experimental} reports a similar phenomenon in grid-tank experiments: ``\textit{The
intensity of these mean fluid motions can be of the order of the particle settling velocity and
therefore strongly affects the measurements.}''

Given the difficulty of measuring both phases simultaneously at our values of $Re_\lambda$, we are unable to asses the impact of these recirculation cells on our results. Future research should address the impact of these weak mean flows on the small Rouse regime. To circumvent these non-zero vertical mean flow effects, we present in section \ref{sc:vall} an analysis in a translating frame of reference.

Considering the experimental difficulties found in the double limit of $\phi_v\to$, and $St\to0$, we will focus our analysis on bulk trends of the moderate Rouse regime, which is less sensitive to these measuring uncertainties.

\section{Moderate Rouse regime}
\subsection{Global behavior}
\label{sc:ganal}
\begin{figure}
	\centering

		\includegraphics[scale=.6]{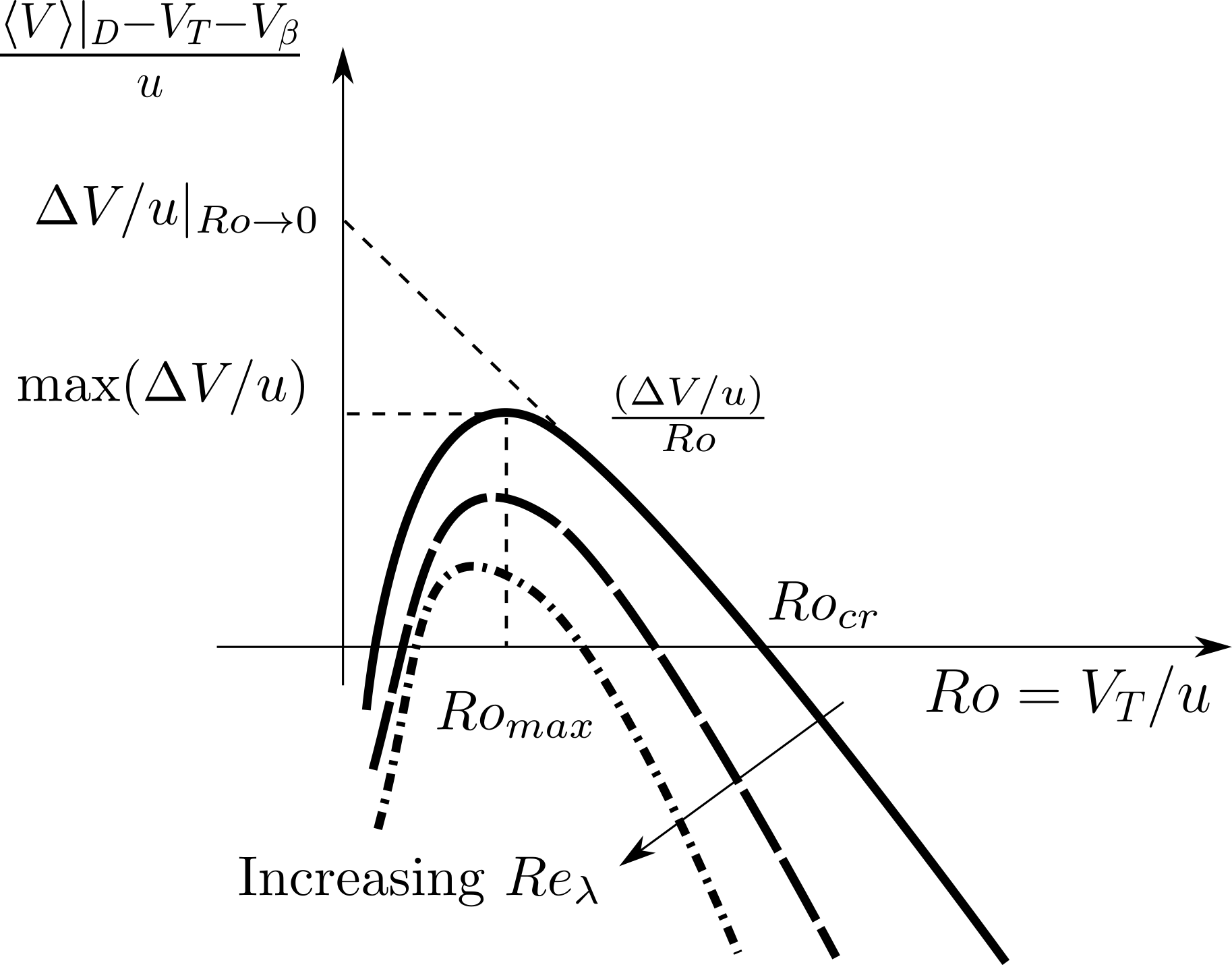}
		\caption{ Parameters computed for the data in \ref{fig-dv-ro} \label{fig-dvst-rosk}. The different line styles refer to different values of $Re_\lambda$.}

\end{figure}

We focus on the regime $Ro>O(0.1)$, and analyze the particles velocity settling curves against Rouse (see figure \ref{fig-dvst-rosk}). The curve is defined by its slope, x-axis intercept (crossover between hindering and enhancement) and its maximum. For those datasets that have not reached hindering, we extrapolated the crossover with a linear fit. 

\begin{figure}
	\centering
	
	\begin{subfigure}[h!]{0.48\textwidth}
	\begin{overpic}[scale=.45]{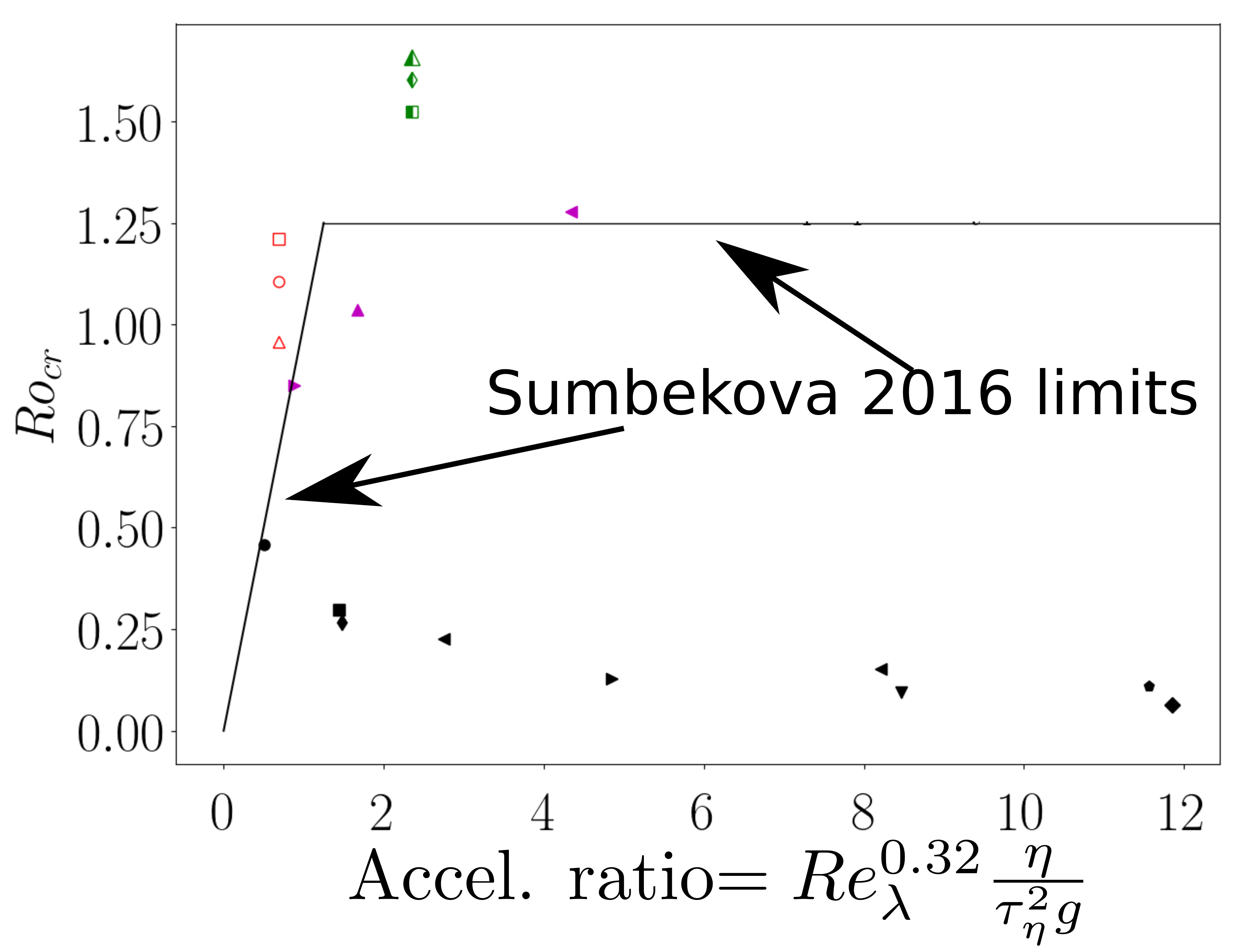}
								\put (85,65) {\huge a)}

	\end{overpic}
	\caption{ \label{fig-dvst-rocr-fra}}
\end{subfigure}
			~
\begin{subfigure}[h!]{0.48\textwidth}
		\begin{overpic}[scale=.45]{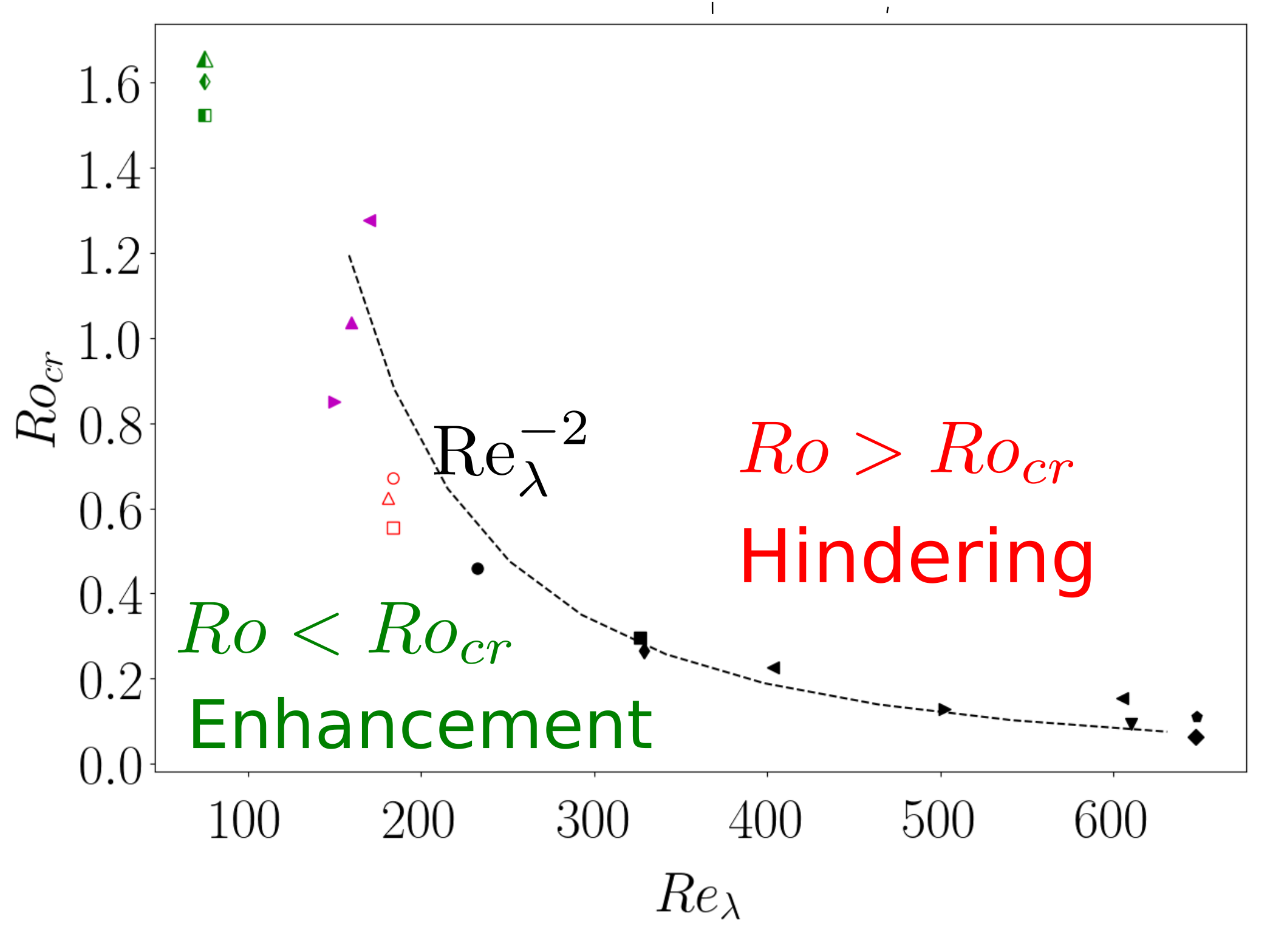}
			\put (85,60) {\huge b)}
	\end{overpic}
	\caption{ \label{fig-dvst-rocr}}
\end{subfigure}

	\begin{subfigure}[h!]{0.48\textwidth}
		\begin{overpic}[scale=.25]{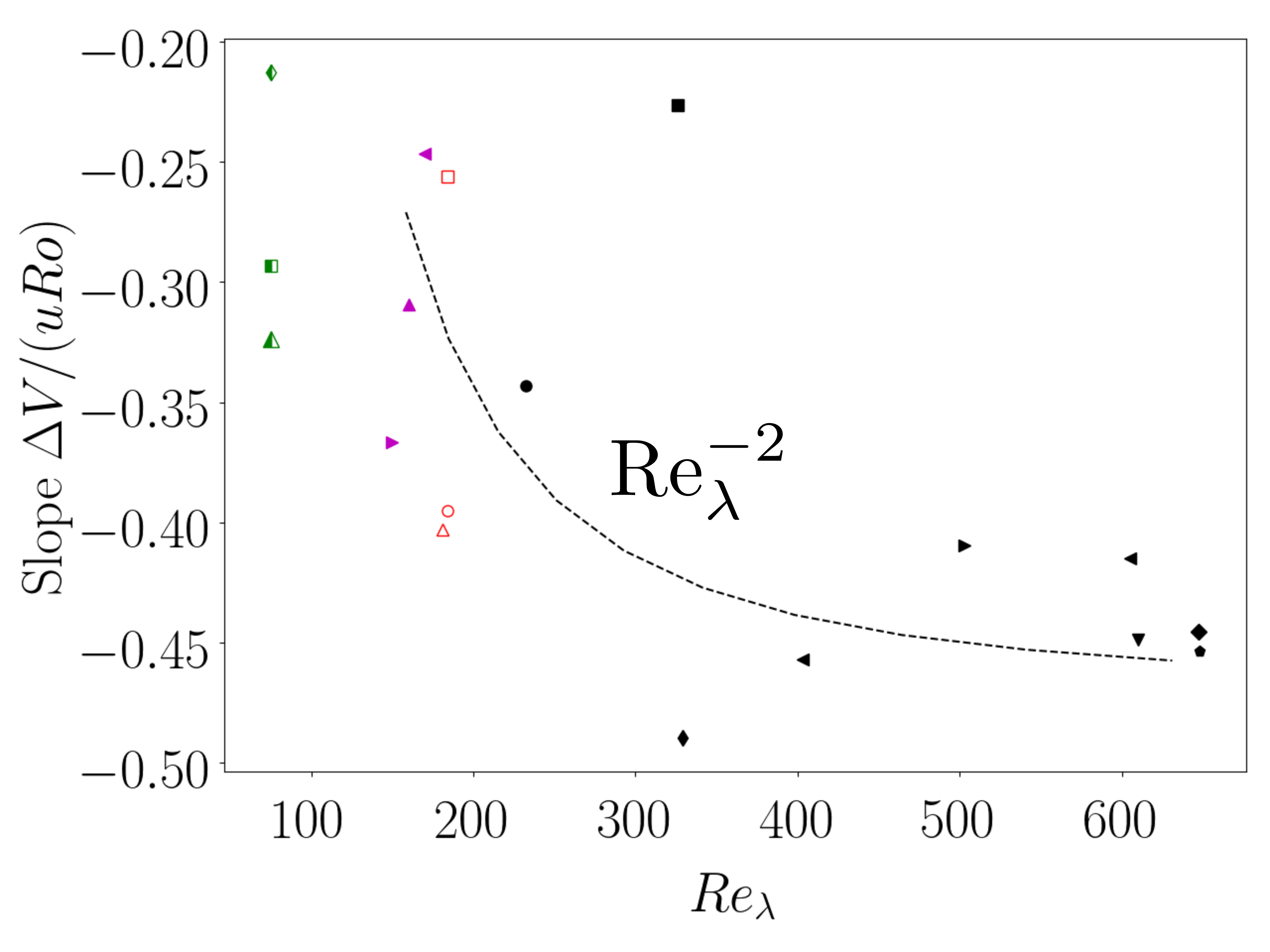}
							\put (85,60) {\huge c)}

\end{overpic}
		\caption{ \label{fig-dvst-slope}}
	\end{subfigure}
			~
	\begin{subfigure}[h!]{0.48\textwidth}
		\begin{overpic}[scale=.25]{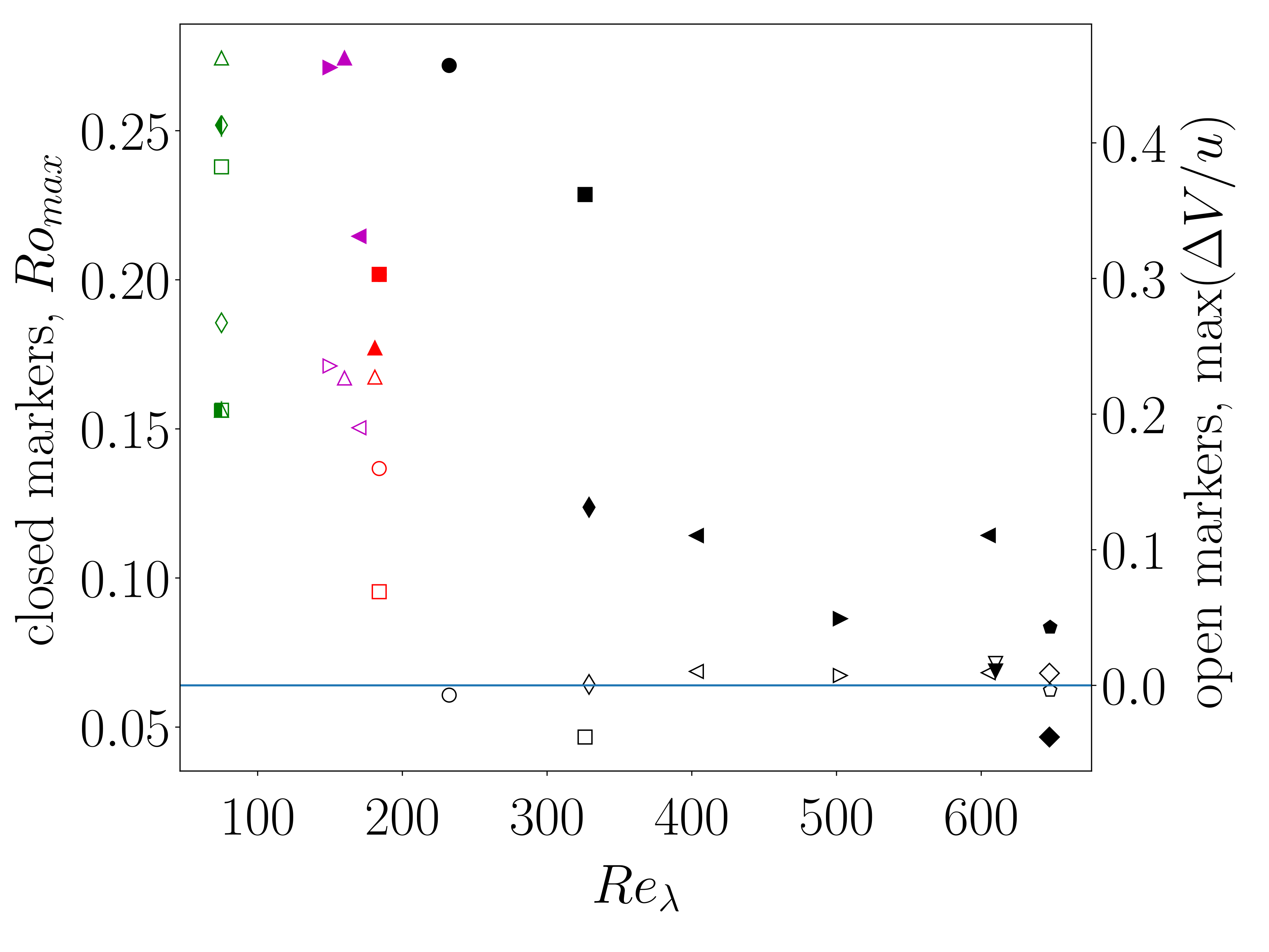}
					\put (75,60) {\huge d)}
					\end{overpic}
		\caption{ \label{fig-peaks}}
	\end{subfigure}

	\vspace{-0.5cm}
	
	\caption{ a) $Ro_{cr}$ cross over between enhancement and hindering against $\gamma_a=\sqrt{0.13Re_\lambda^{0.64}}Fr$. The solid lines refer to the proposed scaling in \cite{Sumbekova2016a}. b) $Ro_{cr}$: crossover value between enhancement and hindering against $Re_\lambda$.  c) Slope of the settling velocity against Rouse number $(\Delta V/u)/Ro$. d) Maximum settling velocity, and Rouse value for these maxima. Markers follow the color convention in figure \ref{fig-dv-ro}.}
	
\end{figure}

First, we consider the scaling of Sumbekova et al. \cite{Sumbekova2019} for a similar range of Rouse numbers (other scalings proposed for this regime are included in Appendix \ref{sc:scalings2}). They propose that the crossover $Ro_{cr}$, which defines the boundary between hindering and enhancement, increases with $\gamma_a=a_0^{1/2}\gamma$, where $a_0=0.13Re_\lambda^{0.64}$ is the Lagrangian acceleration  proposed by Sawford \cite{sawford1991reynolds}. This proposal seems to hold to some extent for previous datasets (see figure \ref{fig-dvst-rocr-fra} ) but it does not for the AG data, which seems to be less affected (if at all) by variations of the fluid acceleration. For the sake of completeness, we also plotted our data using different scalings found in the literature (see appendix \ref{sc:scalings2}).

Interestingly, our data reveal that $Ro_{cr}$ (figure \ref{fig-dvst-rocr}) becomes smaller with increasing $Re_\lambda$, in agreement with \cite{kawanisi2008turbulent,good2012intermittency}. Although the liquid fraction does impact $Ro_{cr}$, the leading order contribution comes from $Re_\lambda$. It is then left for future research to assess whether these effects could be facility dependent (e.g., non-zero mean vertical flow \cite{good2014settling,Sumbekova2016a,akutina2020experimental}).

The linear fit y-intercepts (i.e. the limit $Ro\to0$ in table \ref{tab:sum}) also decrease with increasing $Re_\lambda$. This trend is consistent with the observed reduced settling at increasing $Re_\lambda$ (figure \ref{fig-dvst-rosk}). On the other hand, the fitted linear slopes (figure \ref{fig-dvst-slope}) are of order 1, i.e., $(\Delta V/u)/Ro = \Delta V/V_T=O(1)$, and they seem to become steeper with $Re_\lambda$.  The correlation with $Re_\lambda$, however, is not conclusive, as multiple factors (e.g., recirculation cells, and volume fraction $\phi_v$) could be influencing the results. Interestingly, this quasi-linear behavior has also been recovered in numerical simulations (see appendix \ref{sc:ap1}), where the horizontal motion of the particles was suppressed \cite{rosa2016settling}.

The maximum settling enhancement (figure \ref{fig-peaks}) also decreases with $Re_\lambda$ in agreement with  \cite{sumbekova2016clustering}. Likewise, the Rouse number corresponding to the peak enhancement of the settling velocity decreases with $Re_\lambda$ (figure \ref{fig-peaks}). This observation may be a direct consequence of the coupling between $u$ and $Re_\lambda$ in our experiment: they both scale with the inlet velocity $U_\infty$. Thus, $Ro=V_T/u$ decreases with increasing $Re_\lambda$.  These characteristics of the $\Delta V~vs~Ro$ are summarized in table \ref{tab:sum} in Appendix \ref{sc:ap1}.


\subsection{Local concentration effects}

Some experimental studies report that the increased local concentration has an impact on the particle settling velocity due to preferential concentration \cite{aliseda2002effect,huck2018role}, a mechanism frequently referred to as \textit{collective effects}. Previous research has found, utilizing 2D images, evidence of preferential concentration in the same facility and under the same experimental conditions studied here \cite{sumbekova2017preferential}.

Based on the approach in \cite{obligado2020study}, we decided to normalize $\Delta V$ by the cluster velocity $V_{cl}\sim \langle C_{cl}\rangle \langle A_{cl}\rangle \rho_p g/(\rho_{air}\nu) $, where $\rho_p$ is the particle density, $\langle C_{cl}\rangle$, and $\langle A_{cl}\rangle$ are the clusters concentration, and area, respectively. We estimate the latter quantities as $\langle C_{cl}\rangle\approx 4\phi_v$ from 2D correlations in the same facility \cite{Monchaux2010,sumbekova2017preferential,huck2018role}, and $\langle A_{cl} \rangle=2.1\times 10^{-5}St_{max}^{-0.25}Re_\lambda^ {4.7}\phi_v^{1.2}$ \cite{sumbekova2017preferential}. The mean concentration range has also been reported for anisotropic turbulence \cite{boddapati2020novel} at mass loadings between $1\%$ to $7\%$.

\begin{figure}
	\centering
	\begin{subfigure}[h!]{0.48\textwidth}
	\begin{overpic}[scale=.25]{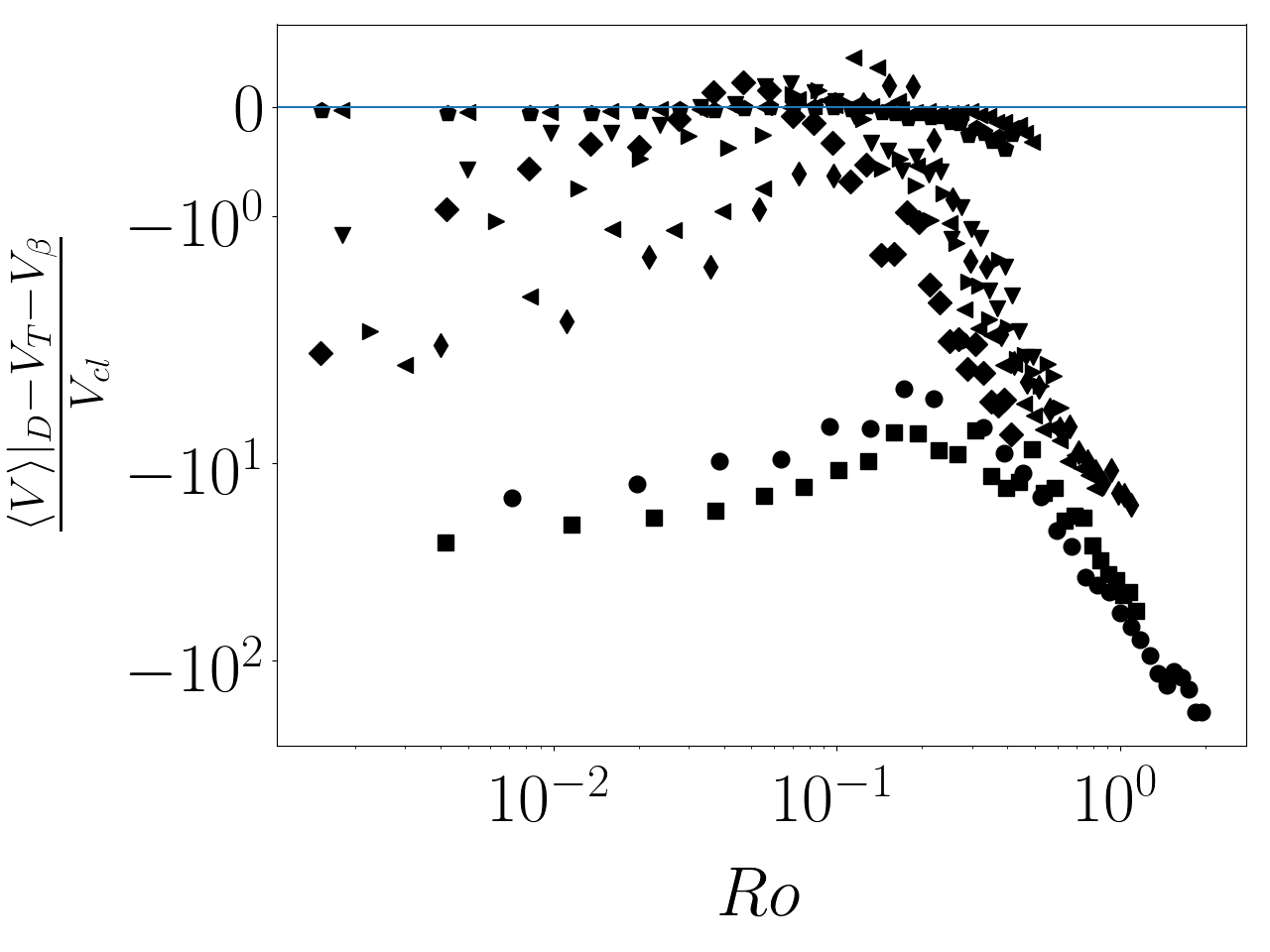}
										\put (30,20) {\huge a)}

		\end{overpic}
		\caption{ \label{fig-dvst-clu}}
	\end{subfigure}
	~
	\begin{subfigure}[h!]{0.48\textwidth}
		\begin{overpic}[scale=.25]{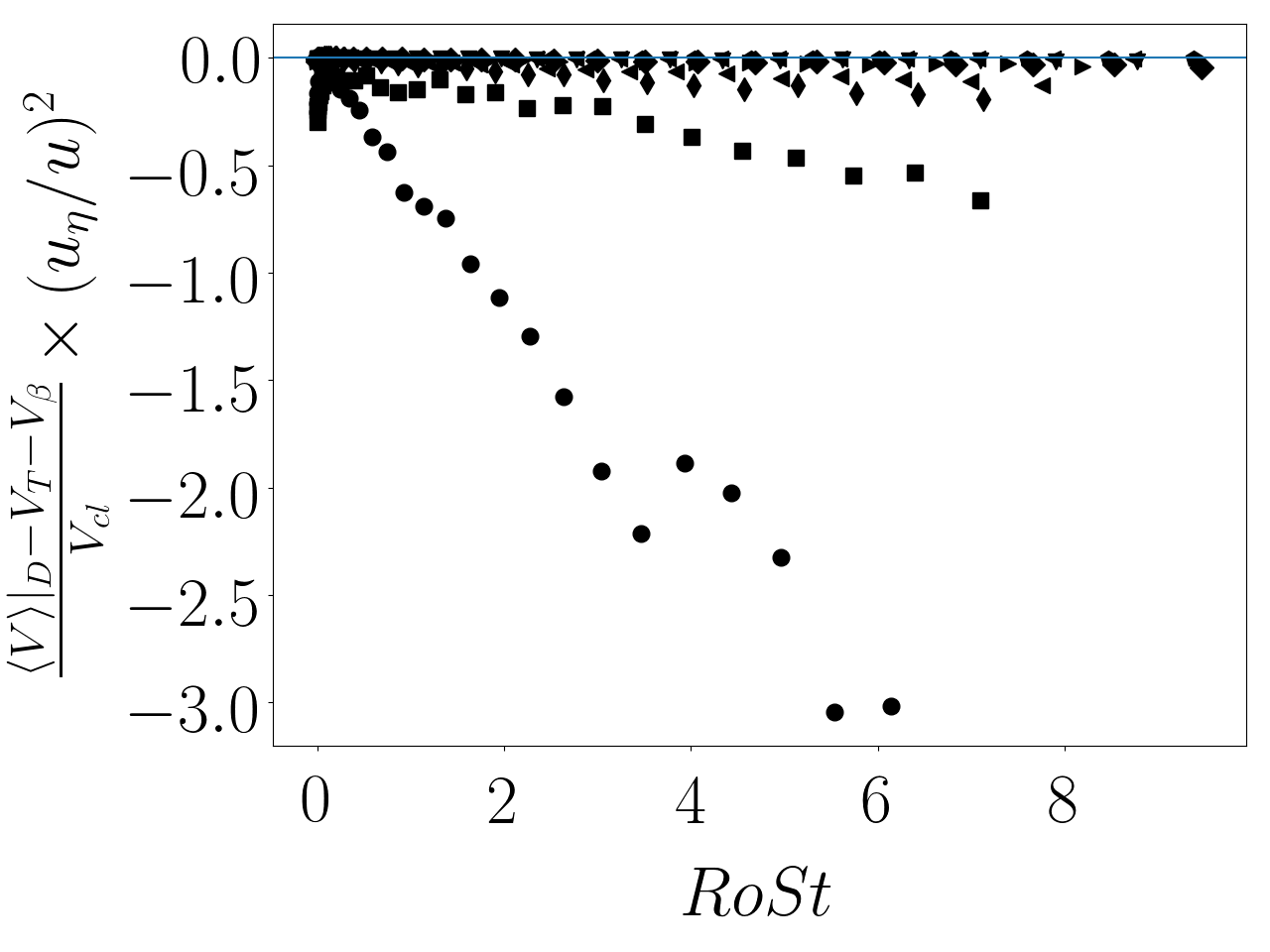}
													\put (30,20) {\huge b)}

		\end{overpic}
		\caption{ \label{fig-dvst-mix-2}}
	\end{subfigure}
	
	\caption{Settling velocity normalized by different scales including the estimated settling velocity $V_{cl}$ following the approach of Obligado et al. \cite{obligado2020study}. The vertical axis in figures a) and b) is negative log, i.e., -1 $\times$ log.  Markers follow the legend found in figure \ref{fig-dv-ro}.}
	
\end{figure}

The normalization by a single velocity scale fails to account for the different trends observed (figure \ref{fig-dvst-clu}). Tom and Bragg \cite{tom2019multiscale} claimed that normalizing the settling velocity results with a single length scale (or velocity scale) may not be adequate due to the multi-scale nature of the turbulence. They advance that the particle settling is affected by the multi-scale phenomenology of turbulent flows, and the resulting particle settling is due to an integrated effect of a range of scales that depend on the particle Stokes number.  Therefore, they argue that some physics may be lost by using a single scale to normalize the particle settling velocity enhancement. 
Tom and Bragg further argue that the multi-scale nature of particle settling explains the better collapse brought by the use of the mixed length scales normalizations (Kolmogorov-scale velocity scaling combined by integral-scale Stokes \cite{good2014settling}). Consistent with their observations, we see a slightly better collapse when using mixed scalings (viscous and integral scales combined) (see figure \ref{fig-dvst-mix-2}).  

\section{Analysis on a moving frame of reference}
\label{sc:vall}

As stated above, non-zero mean vertical flow effects could potentially impact the results presented here. To address these biases, we conduct a final analysis considering the particle settling velocity in a frame of reference moving with the particle distribution global average; $ \langle V\rangle\vert_{all}=\int V(D_p) f(D_p) dD_p$, where $f(D_p)$ is the particle distribution PDF (see figure \ref{fig:spray_pdf}). 

\begin{figure}
	\centering
	\begin{subfigure}[h!]{0.48\textwidth}
		\begin{overpic}[scale=.25]{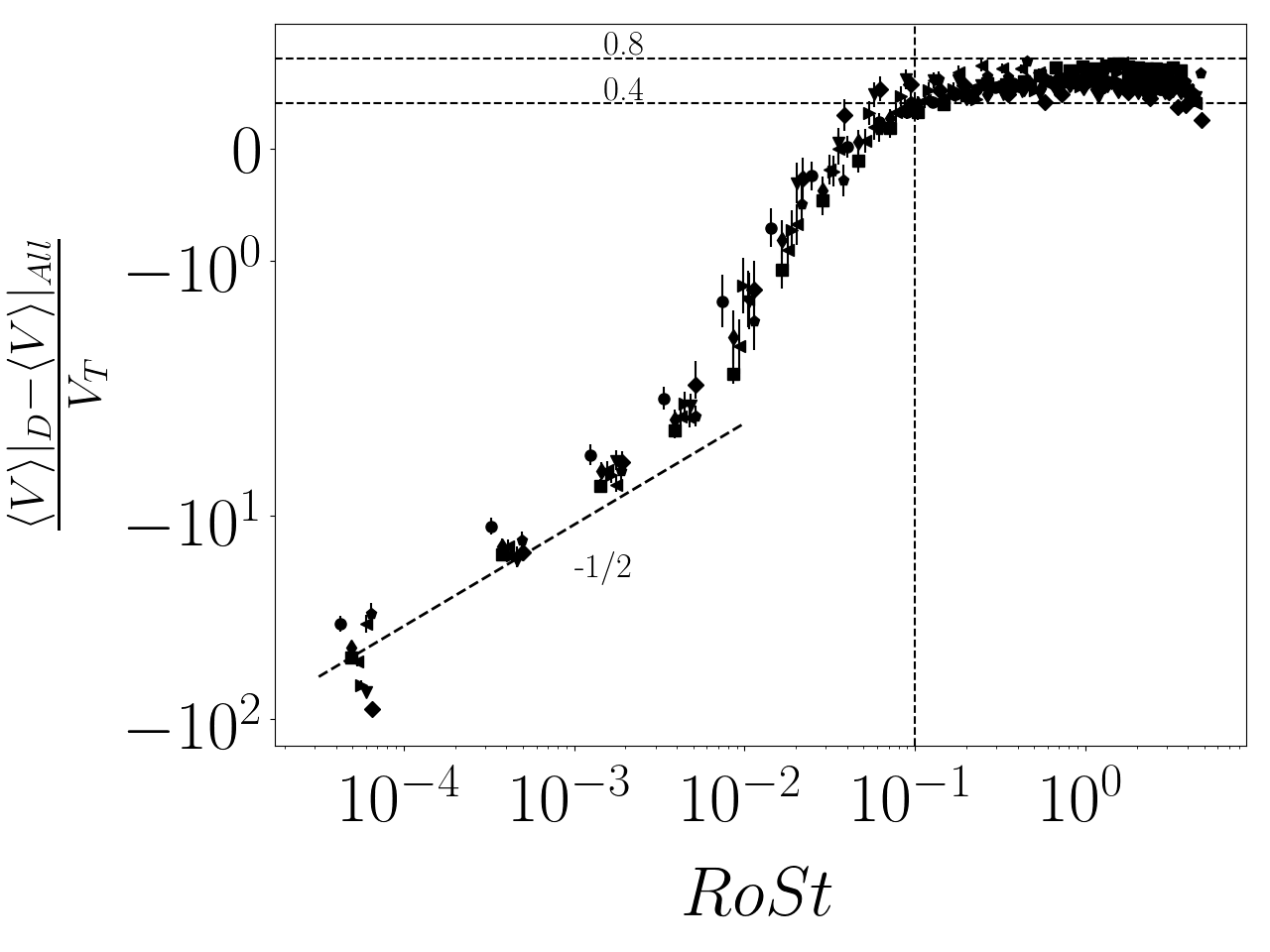}
									\put (85,20) {\huge a)}

		\end{overpic}
		\caption{ \label{fig-vall}}
	\end{subfigure}
	~
	\begin{subfigure}[h!]{0.48\textwidth}
		\begin{overpic}[scale=.25]{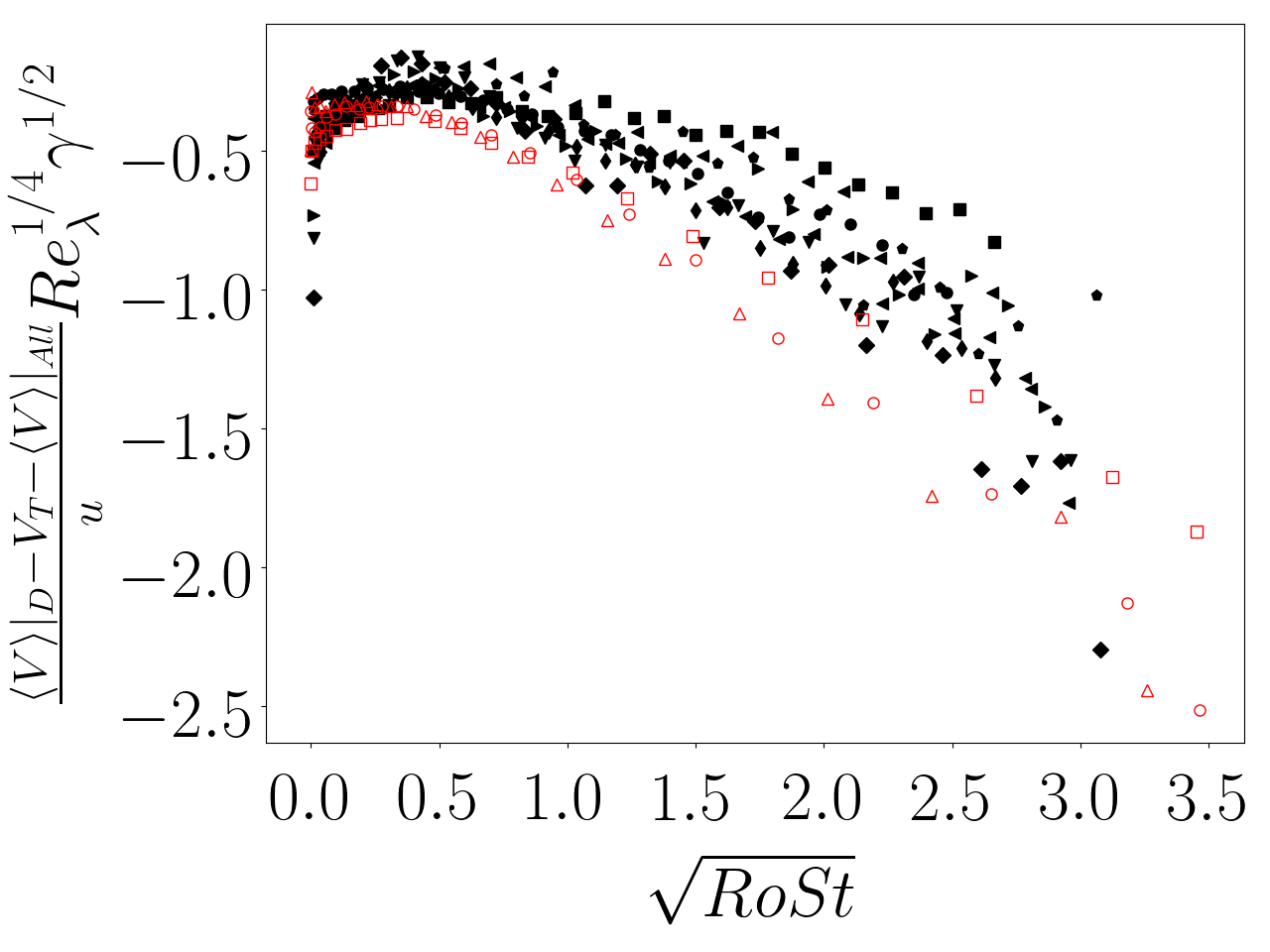}
											\put (30,20) {\huge b)}

		\end{overpic}
		\caption{ \label{fig-vall-2}}
	\end{subfigure}

	\caption{a) Settling Velocity in a relative frame. Error bars account for the velocity vertical resolution $\pm0.005ms^{-1}$. b) Scaling of equation  \ref{eq:RosTS} applied in the relative moving frame of reference.  Markers follow the legend found in figure \ref{fig-dv-ro}. }
	
\end{figure}

In this moving frame of reference, we encounter the following question: \textit{which parameter does control the evolution of the particle in the translating frame of reference?} After some iteration, we find that the scaling $RoSt$, combining the Rouse and the Stokes numbers (see equation \ref{eq:Rost}), provides the best collapse of the data (see figure \ref{fig-vall}) in the x-coordinate that $Ro$ or $St$ individually.

Interestingly, in the regime $RoSt>0.1$, the relative particle settling velocity has a slow evolution (see figure \ref{fig-vall}): 

\begin{equation}
\frac{\langle V \rangle\vert_{D}-\langle V \rangle\vert_{all}}{V_T}\approx C ,
\label{eq:vvuc}
\end{equation}

with $C\in [0.4-0.5]$, and which after algebraic manipulation gives;

\begin{equation}
\frac{\langle V \rangle\vert_{D}-\langle V \rangle\vert_{all} - V_T}{u}\approx (C-1) Ro.
\label{eq:rvro}
\end{equation}

This expression is consistent with the quasi-linear behavior found in figure \ref{fig-dv-ro}. Although the datasets present some variability at small Rouse numbers, we observe a power-law dependency for small $RoSt\ll 10^{-2}$. If we were to apply this observed power law, algebraic manipulations would yield:

\begin{equation}
\frac{\langle V \rangle \vert_{D}-V_T-\langle V \rangle\vert_{all}}{u}\approx C_\dagger\Big(\frac{15^{1/4}}{\gamma Re_\lambda^{1/2}}\Big)^{1/2}-Ro.
\label{eq:RosTS}
\end{equation}

This result suggests that at very small Rouse numbers it would be possible to bound these profiles within the values of parameter $C_\dagger$. The data has a better collapse in this framework when premultiplied by the mixed scaling (see figure \ref{fig-vall-2}). The effects of 
$\langle V\rangle\vert_{all}$ and its relationship with the particle size distribution and the observed particle settling should be further investigated in future experiments. For instance, some experiments have advanced that a bidisperse particle distribution may fall faster than any of the two monodisperse ones \cite{wang2020clustering}, an enhancement that cannot be explained by simple linear superposition, i.e., by taking an effective diameter of the bidisperse distribution.

\section{Concluding remarks}

 Using phase Doppler interferometry, we experimentally investigate the behavior of polydipersed inertial sub-Kolmogorov particles under homogeneous isotropic turbulence for turbulent Reynolds numbers up to $\approx 650$. Combined with previously available experimental results in the range $Re_\lambda\in [75-200]$ taken in different facilities \cite{aliseda2002effect,good2014settling,Sumbekova2019}, we find that the average settling velocity of particles is mainly a function of the Rouse number of the particle ($Ro$) and the overall particle-turbulence interactions are governed by the Taylor-based Reynolds number ($Re_\lambda$). Other parameters such as the ratio between  the rms acceleration and gravity ($\gamma$) seem to have a very small (if at any) contribution to the particle settling behavior.
 
Our results also suggest that at increasing values of $Re_\lambda$, the particles settling velocity is increasingly hindered: their measured particle settling velocity is smaller than their respective one in still fluid. This observation is recovered for all particle sizes (and therefore Stokes and Rouse numbers) for each experimental condition explored here.

Close inspection of this difference between the measured particle settling velocity and their respective one in still fluid, reveals that the boundary between the particle settling hindering ($\Delta V/V_T<1$) and enhancement  ($\Delta V/V_T>1$) regimes depends on $Re_\lambda$. The onset of such transition point seems to behave as $Re_\lambda^{-2}$.  In addition, after the peak of enhancement, $\Delta V / u^{\prime} \approx -\kappa Ro$ decreases almost linearly with the Rouse number. This behavior starts in the enhancement region and goes well into the hindering region for all the Rouse numbers considered here, $Ro<10$. Noteworthy,  the $\kappa$ parameter, which accounts for this linear behavior, seems also to be a function of $Re_\lambda$ for a fixed particle distribution: the larger the $Re_\lambda$, the steeper the slope. 

 Although our concentration range is narrow to reach a definite conclusion, we do not recover a strong influence of the concentration on the results presented. This lack of influence seems to be a consequence of the more dilute conditions of our experiments ($\phi_v\leq O(10)^{-5}$) with respect to those conducted in the same facility \cite{sumbekova2017preferential}, which report the existence of preferential concentration. Previous studies \cite{aliseda2002effect,huck2018role} have shown that the existence of preferential concentration leads to enhanced settling velocity for those particles inside a clusters. Sumbekova et al. \cite{sumbekova2017preferential} reports that the degree of clustering, as well as the clusters' characteristic size is an increasing function of $Re_\lambda$. On the contrary, we conjecture that these \textit{collective effects} \cite{huck2018role} become less important at increasing values of $Re_\lambda$ where the hindering effect takes control of this phenomenon. 2D PTV measurements taken in the same facility support such conjecture: \cite{sumbekova2016clustering} (figure 4) reports that for a fixed droplet distribution, increasing $Re_\lambda$ leads to a global reduction in the measured particle settling velocity for particles inside clusters.

Finally, we cannot rule out that our wind tunnel experiments might be affected by a non-zero mean vertical velocity, as proposed by previous research \cite{good2014settling,Sumbekova2016a}. To address this potential bias, we have plotted our data in a translating frame of reference moving at the mean vertical velocity of our particle distribution. Previous experimental data as well as ours seem to collapse better in this frame, and it aids in explaining the quasi-linear behavior in the absolute (laboratory) frame of reference past the peak enhancement.

\section{Acknowledgements}

Our work has been partially supported by the LabEx Tec21 (Investissements d'Avenir - Grant Agreement $\#$ ANR-11-LABX-0030), and by the ANR project ANR-15-IDEX-02. We also thank Laure Vignal and Vincent Govart for their help with the experiments.

\appendix
\section{Experimental setup details}
\label{sca:exps}
\subsection{Droplet injection specifics}

\begin{figure}
	\centering
	\begin{center}	
		
	\begin{subfigure}[t]{0.48\textwidth}
			\begin{overpic}[scale=0.8]{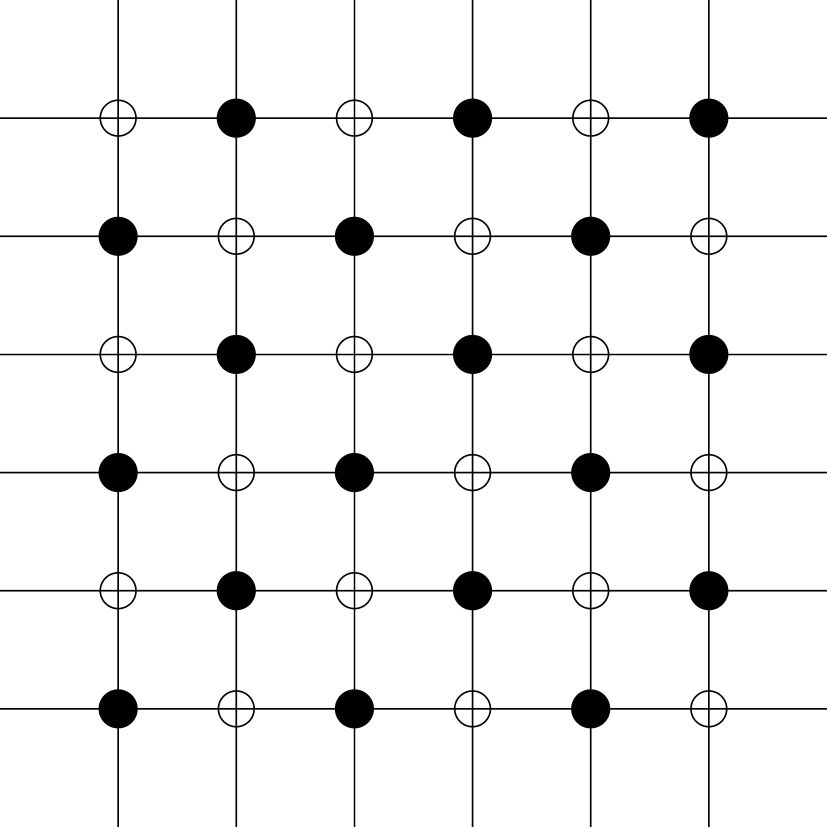}
				\put (0,5) {\huge a)}

			\end{overpic}
			\caption{\label{fig:injr}}
		\end{subfigure}
		~
 		\begin{subfigure}[t]{0.48\textwidth}
			\begin{overpic}[scale=0.5]{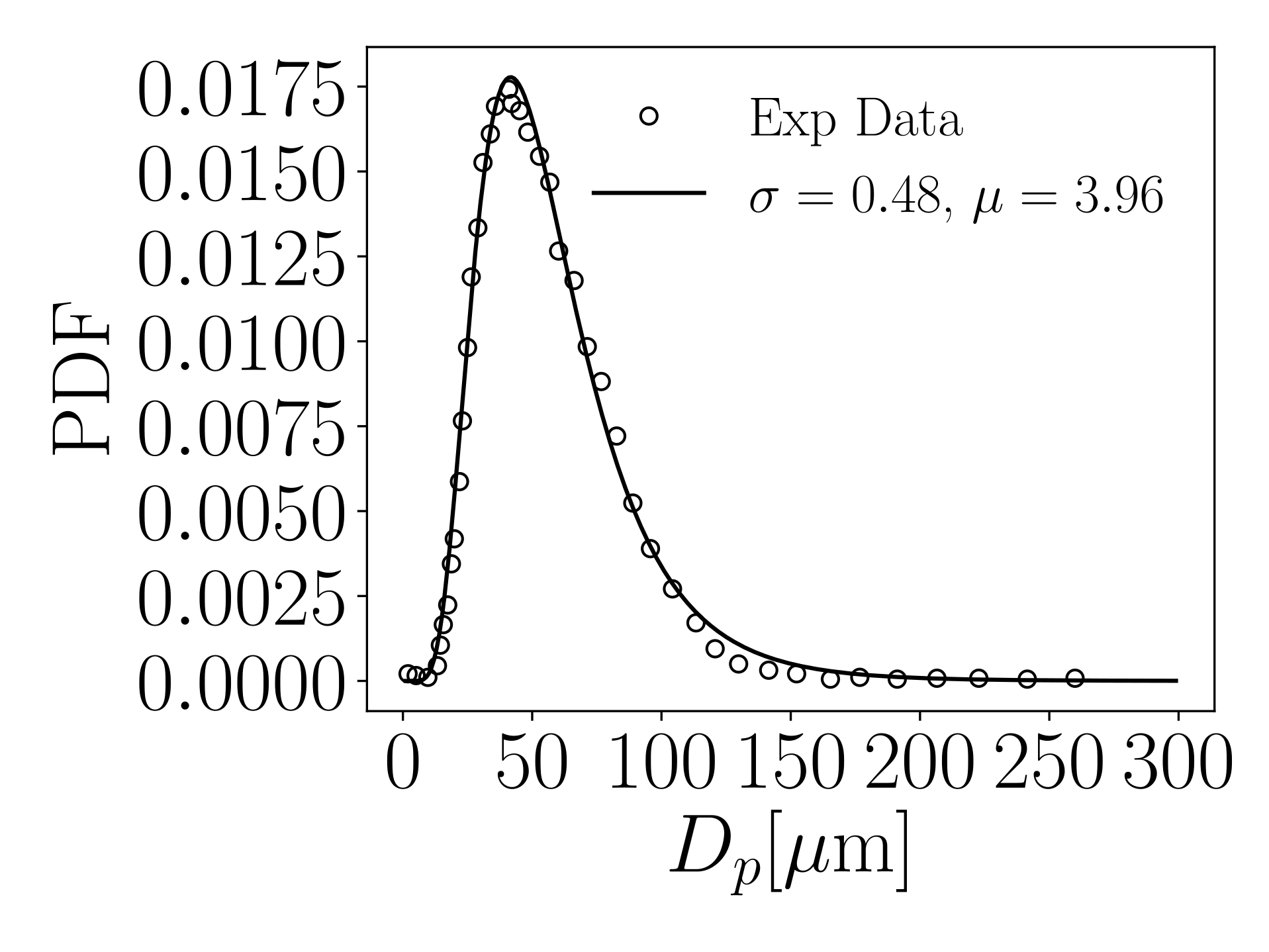}
							\put (5,5) {\huge b)}

			\end{overpic}
			\caption{\label{fig:spray_pdf}}
		\end{subfigure}
	\end{center}

	\caption{a) Injector rack sketch. For the lowest volume fractions half of the injectors (filled markers) were utilized. b) Spray characterization coming from PDI data from Sumbekova \cite{Sumbekova2016a}. }	
\end{figure}

\begin{figure}[t]
\centering
			\includegraphics[scale=1.2]{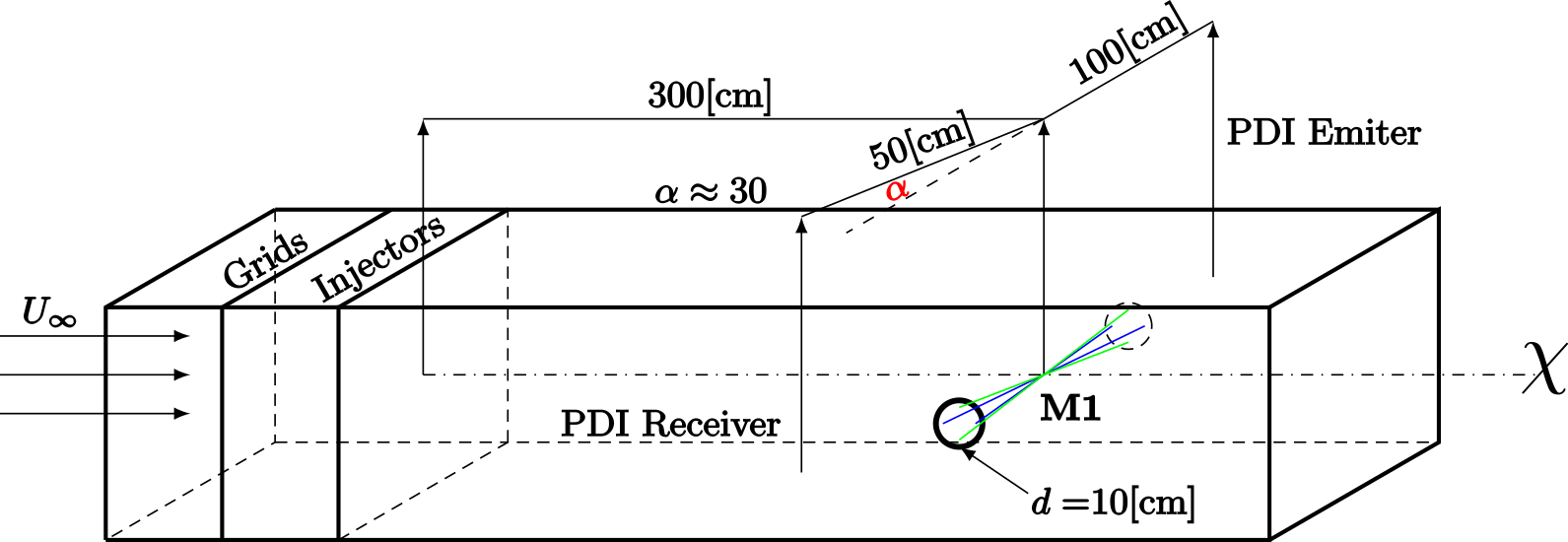}

\caption{Sketch of our experimental setup (not to scale). The wind-tunnel cross-section is 75x75 cm$^2$. Its center line is labeled as $\chi$ in the figure. The emitter and receiver components of the PDI are on the same horizontal plane. However, the receiver is positioned at 30 degrees (see $\alpha$ in the figure) with respect to the emitter to maximize the capture of the water droplets refraction \cite{bachalo1984phase,Sumbekova2019}. Two holes of approximately 10 cm were carved onto the walls to counteract the water accumulation on them. The measuring station was located at the position labeled as \textbf{M1} on the wind-tunnel center line, and 3 meters downstream of the droplets injection.\label{fig:WT_sk} }
\end{figure}

Downstream of the `grid' section (see figure \ref{fig:WT_sk}) a rack of 18, or 36 spray nozzles --at smaller concentrations fewer injectors were used, see figure \ref{fig:injr}-- injected inertial water droplets with diameters $D_p$ between 20 and 300 microns , i.e, $D_p\in[20-300] \,\mu$m.  This polydispersity was measured by phase Doppler interferometry (PDI). The droplet distribution is close to log-normal distribution (see fit in figure \ref{fig:spray_pdf}). The droplets were considered as spherical particles as their Weber number parameter was, for most droplets, below unity (see in Sumbekova \cite{Sumbekova2016a} section 6.3).

The measuring station was placed 3m downstream of the droplet injection (see figure  \ref{fig:WT_sk}).  The measuring volume lies at the centerline of the wind-tunnel. We used a PDI (Artium-PDI-200) apparatus, which can measure the settling velocity and the particles' diameter simultaneously \cite{bachalo1984phase,schwarzkopf2011multiphase}. The PDI setup has two components: the receiver and the laser emitter. The laser emitter was placed perpendicular to the gas flow. The receiver (see figure \ref{fig:WT_sk}) was on the same horizontal plane but rotated 30 degrees to ensure adequate capture of spherical water droplets in the airflow.  

To quantify the effects of the carrier phase turbulence on the particles, we tried to match as close as possible, the particles' mean horizontal velocity ($\langle U_p \rangle$) to respective unladen mean velocity ($U_\infty$), i.e., in our experiments $\langle U_p \rangle\approx U_\infty$. The validation rate (valid droplet measurements over the total number of droplets detected) reported by the PDI software was above 70$\%$ or higher in all experimental realizations. The acquisition rate  (particles per second) varied between 400 and 3000 Hz depending on the liquid fraction, and bulk velocity, i.e., a higher concentration at a higher bulk velocity gave a higher acquisition rate.

\subsection{Particle velocity PDFs}

Particle velocity PDFs supporting the claim of Gaussian statitics made on section \ref{sc:exps}.

 \begin{figure}
	
	\begin{center}
		\begin{subfigure}[t]{0.48\textwidth}
			\begin{overpic}[scale=0.17]{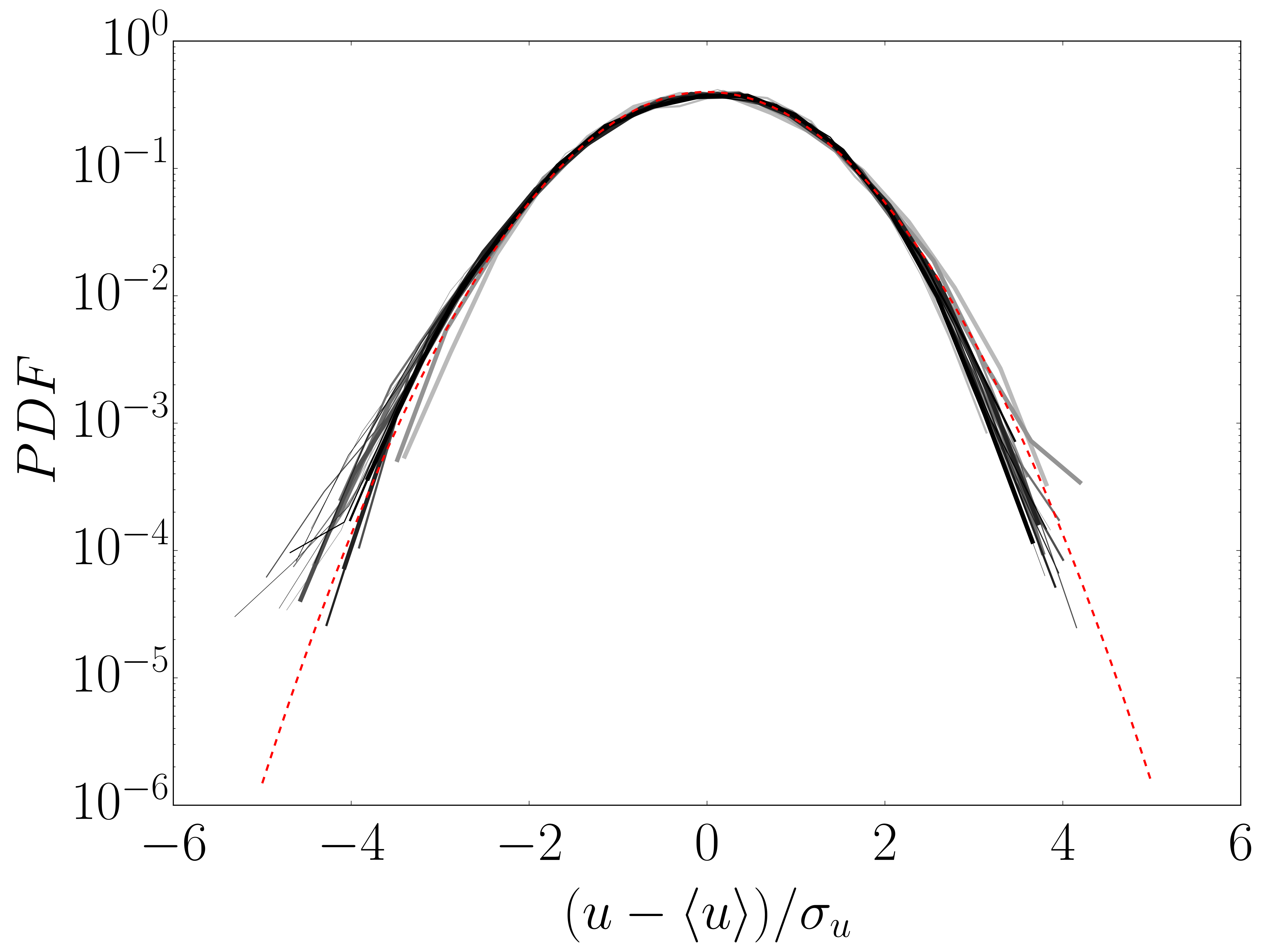}
			\put (85,60) {\huge a)}
            \end{overpic}
			\caption{\label{fig:PDFUAG}}
		\end{subfigure}
		~
		\begin{subfigure}[t]{0.48\textwidth}
			\begin{overpic}[scale=0.17]{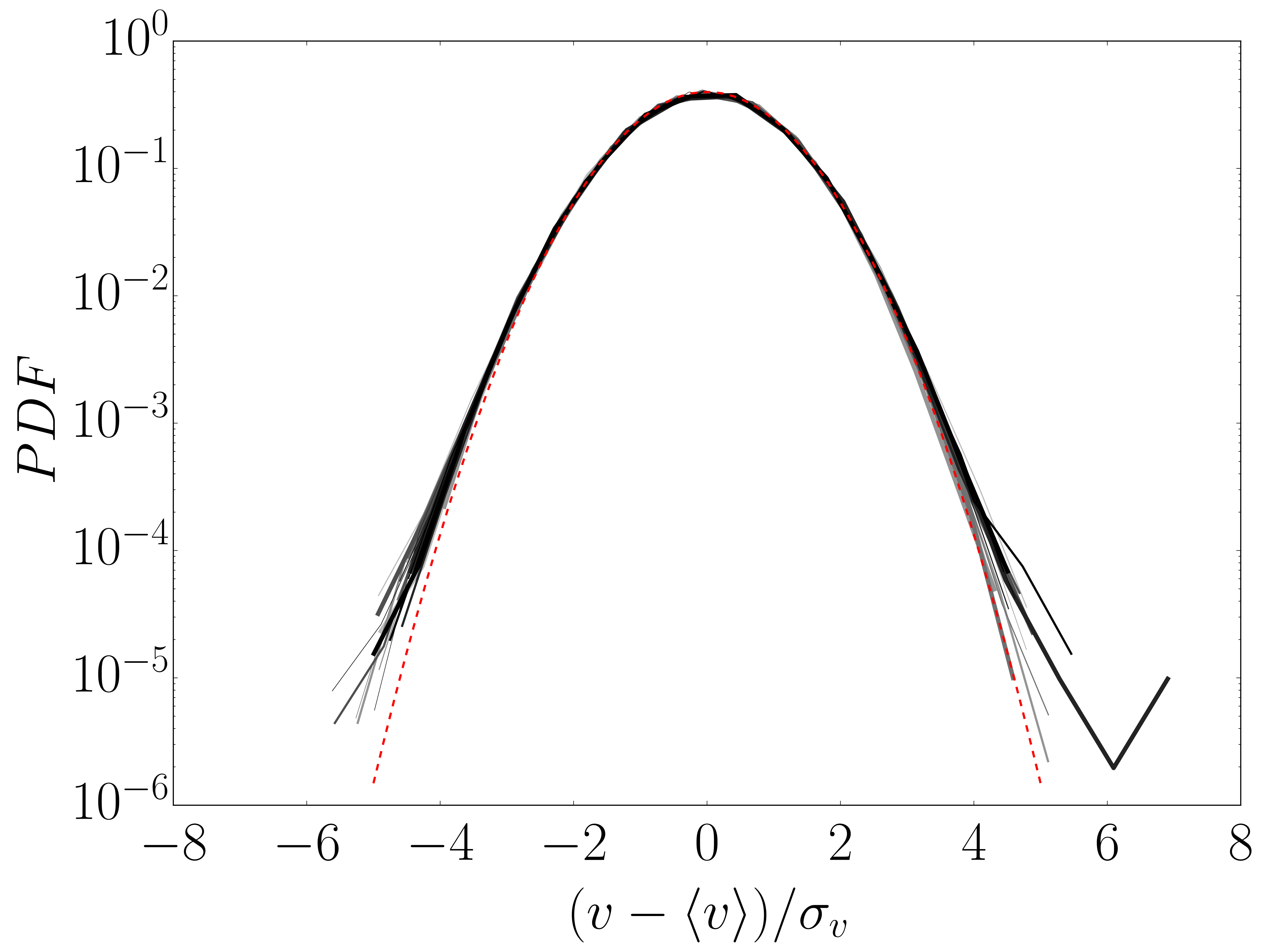}
						\put (85,60) {\huge b)}

			\end{overpic}
			\caption{\label{fig:PDFVAG}}
		\end{subfigure}
	\end{center}

	\caption{ PDFs of the particles velocity for the different records. a) Horizontal component. b) Vertical component.
 The darker the color the larger $Re_\lambda$. In the figures, the normal distribution is plotted as a dashed line (\dashed).\label{fig:velpdf}}  
\end{figure}

\section{Global settling velocity behavior against Rouse number trends}
\label{sc:ap1}
Table \ref{tab:sum} summarizes the different parameters collected from the analysis made in section \ref{sc:ganal}. 

\begin{table}
	\begin{tabular}{l|c|c|c|c|c|c|c|c|c|c|c}
		
		{} &  $10^5\phi_v$ & $Re_\lambda$ &   $\gamma$ & $\varepsilon$ & $\eta$ &   Slope & $\Delta V/u\vert_{Ro\to0}$ & $Ro_{cr}$ & $Ro_{max}$ & $(\mathrm{max}(\Delta V/u)$ \\
		AEA 1  &           1.5 &           75 &  1.630 &         1.000 &    241 &  -0.213 &                      0.341 &     1.602 &      0.252 &                       0.267 \\
		AEA 2  &           6.0 &           75 &  1.630 &         1.000 &    241 &  -0.293 &                      0.446 &     1.523 &      0.156 &                       0.382 \\
		AEA 3  &           7.0 &           75 &  1.630 &         1.000 &    241 &  -0.324 &                      0.536 &     1.657 &      0.156 &                       0.463 \\
		GEA E1 &           0.1 &          150 &  0.500 &         0.200 &    360 &  -0.367 &                      0.312 &     0.851 &      0.215 &                       0.190 \\
		GEA E2 &           0.1 &          160 &  0.900 &         0.460 &    290 &  -0.309 &                      0.321 &     1.037 &      0.274 &                       0.227 \\
		GEA E3 &           0.1 &          170 &  2.300 &         1.600 &    220 &  -0.247 &                      0.315 &     1.277 &      0.271 &                       0.236 \\
		SBK 1  &           0.5 &          185 &  0.490 &         0.200 &    400 &  -0.256 &                      0.310 &     0.555 &      0.202 &                       0.069 \\
		SBK 2  &           1.0 &          185 &  0.490 &         0.200 &    400 &  -0.395 &                      0.436 &     0.671 &      0.202 &                       0.160 \\
		SBK 3  &           2.0 &          185 &  0.490 &         0.200 &    400 &  -0.405 &                      0.386 &     0.624 &      0.177 &                       0.227 \\
		This study     &           0.9 &          232 &  0.243 &         0.078 &    455 &  -0.343 &                      0.157 &     0.459 &      0.272 &                      -0.007 \\
		This study     &           0.6 &          326 &  0.625 &         0.277 &    332 &  -0.226 &                      0.067 &     0.297 &      0.229 &                      -0.038 \\
		This study     &           1.0 &          329 &  0.641 &         0.286 &    330 &  -0.490 &                      0.130 &     0.266 &      0.124 &                       0.001 \\
		This study     &           0.7 &          403 &  1.118 &         0.601 &    274 &  -0.457 &                      0.104 &     0.227 &      0.114 &                       0.010 \\
		This study     &           0.6 &          503 &  1.840 &         1.168 &    232 &  -0.410 &                      0.052 &     0.128 &      0.086 &                       0.007 \\
		This study     &           0.5 &          610 &  3.014 &         2.255 &    197 &  -0.449 &                      0.042 &     0.094 &      0.069 &                       0.016 \\
		This study     &           1.0 &          605 &  2.934 &         2.176 &    198 &  -0.415 &                      0.064 &     0.153 &      0.114 &                       0.009 \\
		This study     &           0.4 &          647 &  4.141 &         3.444 &    177 &  -0.445 &                      0.028 &     0.063 &      0.047 &                       0.009 \\
		This study     &           0.8 &          648 &  4.040 &         3.333 &    178 &  -0.454 &                      0.050 &     0.110 &      0.083 &                      -0.004 \\
	\end{tabular}
	
	\caption{Summary of the parameters extracted from figures \ref{fig-dvst-rocr} to \ref{fig-peaks}. \label{tab:sum}}
\end{table}

\section{Alternative scalings}
\subsection{Normalization by $V_T$}
\label{sc:scvt}

\begin{figure}
	\centering
	\begin{subfigure}[h!]{0.48\textwidth}
				\begin{overpic}[scale=.25]{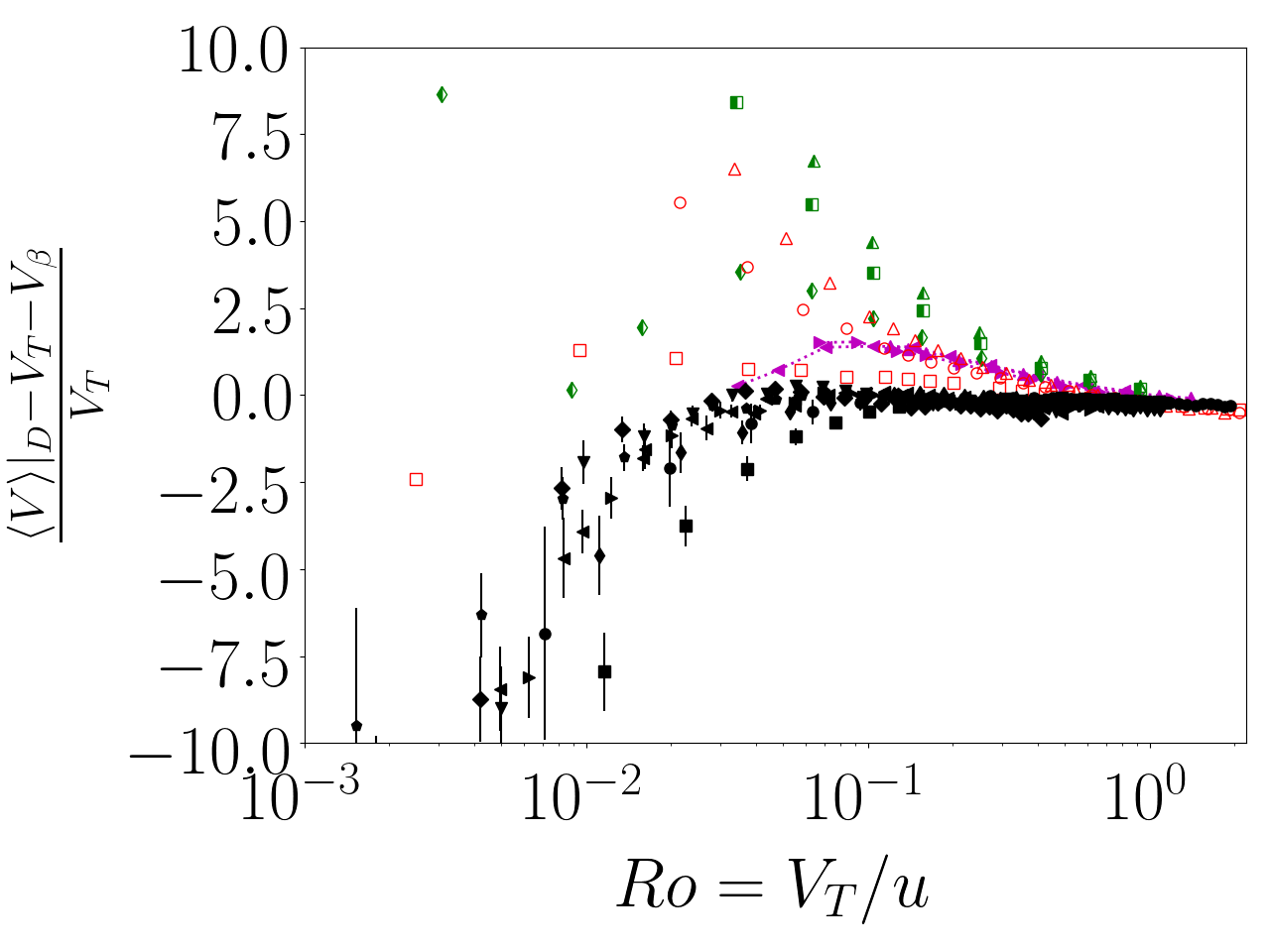}
	\put (85,60) {\huge a)}
        \end{overpic}
		\caption{ \label{fig-dvst-ro}}
	\end{subfigure}
~
	\begin{subfigure}[h!]{0.48\textwidth}
	\begin{overpic}[scale=.25]{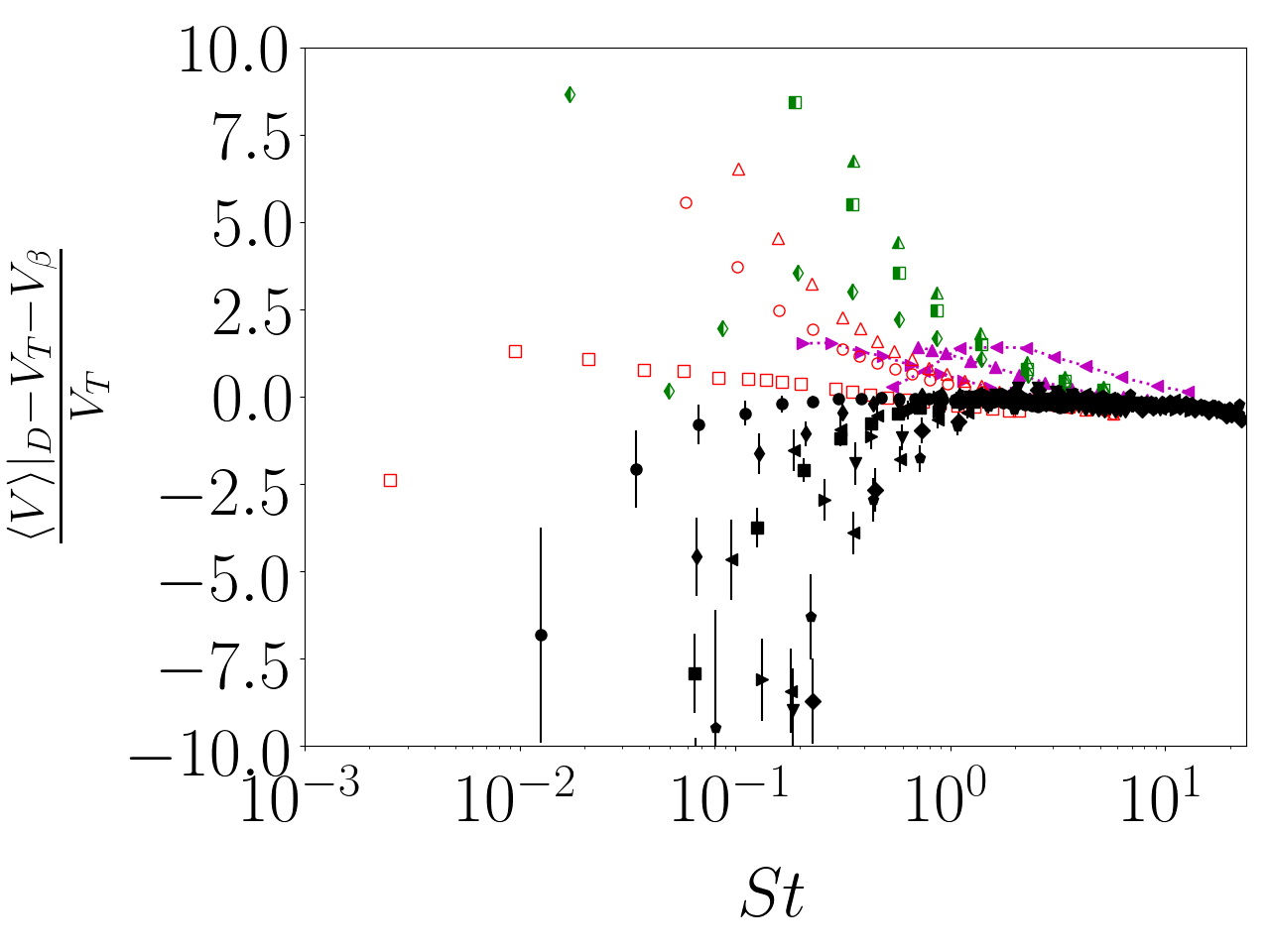}
		\put (85,60) {\huge b)}

		        \end{overpic}

		\caption{ \label{fig-dvst-st}}
	\end{subfigure}

	\caption{Particle velocity over the particle terminal speed. a) Against Rouse. b) Against stokes. The markers follow the legend of figure \ref{fig-dv-ro}.}
	
\end{figure}

If the particles datasets were to be normalized by the respective terminal velocity, we obtain the results found in figures \ref{fig-dvst-ro} and  \ref{fig-dvst-st}.

\subsection{Sumbekova et al. \cite{Sumbekova2019}}
\begin{figure}
	\centering
	\begin{subfigure}[h!]{0.48\textwidth}
		\begin{overpic}[scale=.25]{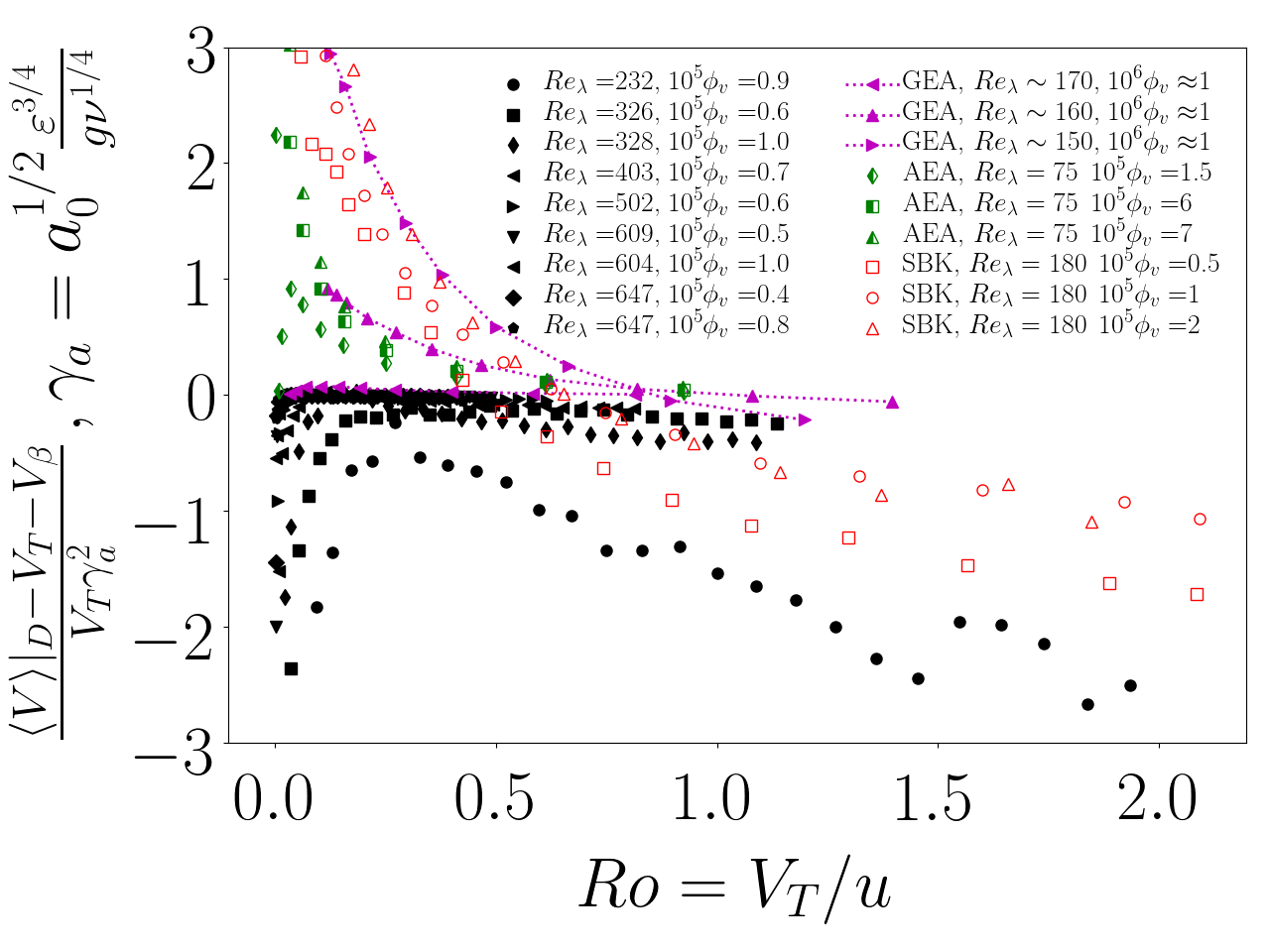}
								\put (30,20) {\huge a)}

		\end{overpic}
		\caption{ \label{fig-app-1}}
	\end{subfigure}
	~
	\begin{subfigure}[h!]{0.48\textwidth}
		\begin{overpic}[scale=.25]{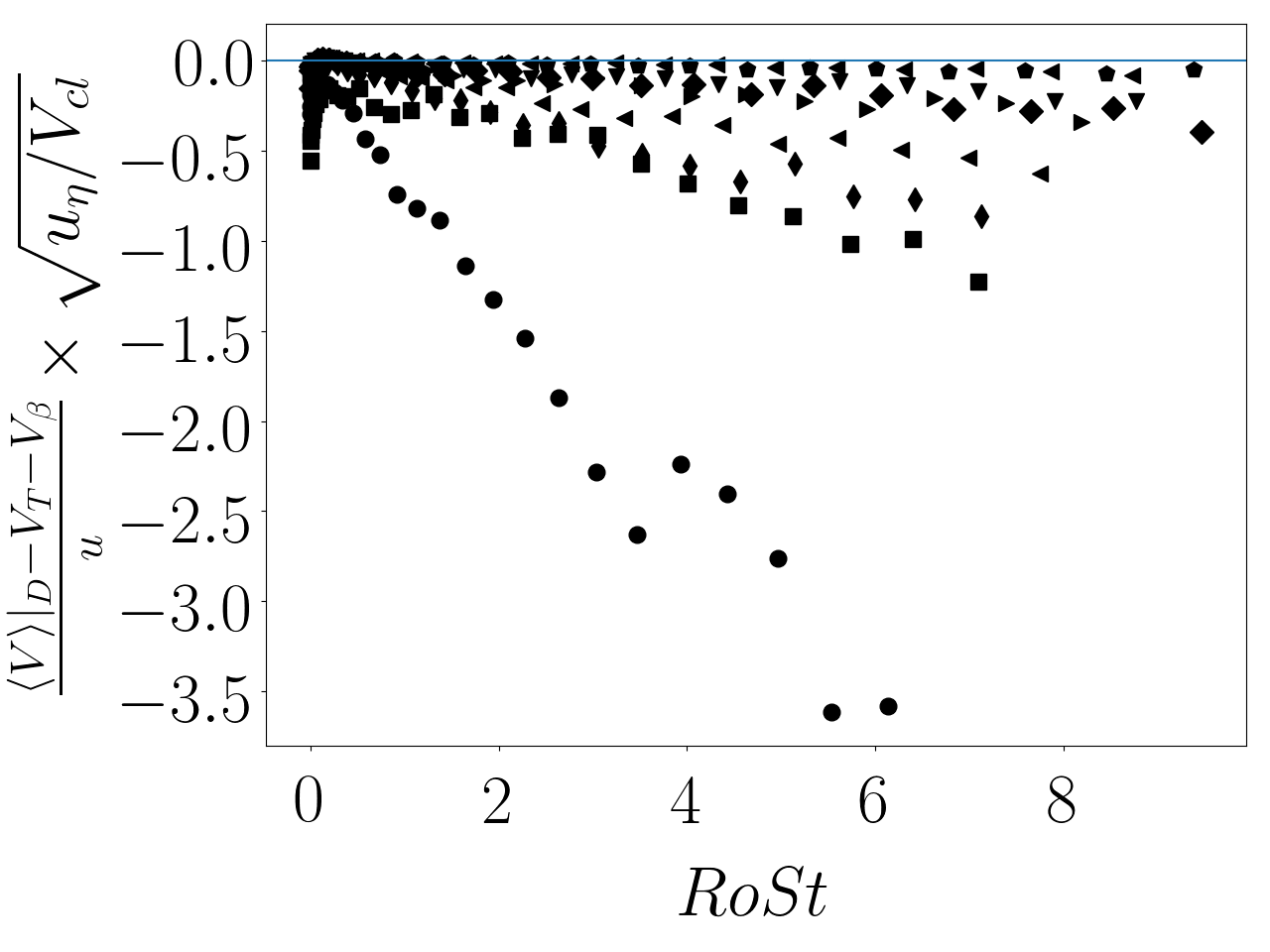}
						\put (85,20) {\huge b)}

		\end{overpic}
		\caption{ \label{fig-app-2}}
	\end{subfigure}
	
	\begin{subfigure}[h!]{0.48\textwidth}
	\begin{overpic}[scale=.5]{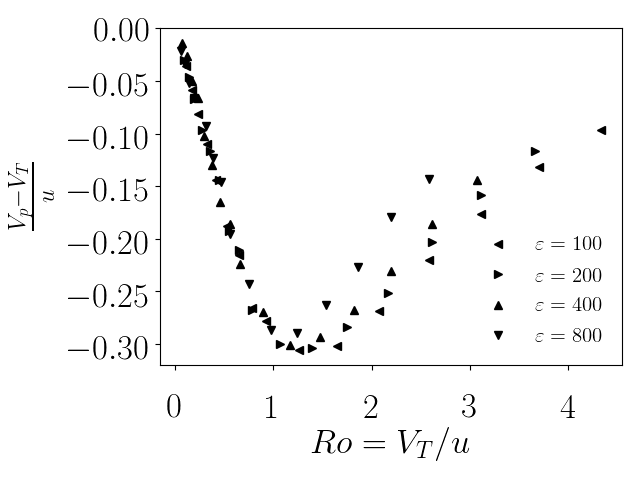}
				\put (85,60) {\huge c)}

		\end{overpic}
		\caption{ \label{fig-app-3}}
	\end{subfigure}

	\caption{a) Sumbekova et al. scaling \cite{Sumbekova2019}. b) Combination of the velocity scales for the AG data. c) Data from fig 16 of Rosa et al. \cite{rosa2016settling}. In the legends, GEA the data of Good et al.\cite{good2014settling}, AEA refers to the data of Aliseda et al. \cite{aliseda2002effect}, and SBK refers to the data of Sumbekova \cite{Sumbekova2019}. }
	
\end{figure}

The scaling of Sumbekova et al. \cite{Sumbekova2019} (figure \ref{fig-app-1}) does not show a better collapse when compared to those of include in the main text. In the figure, some of the curves look closer, but this could be an effect of the y scale used. On the other hand, when large and small fluid scales are are combined with the cluster falling velocity the curves come close together to some extent (figure \ref{fig-app-2}). This highlights again the including multiple scales may be necessary to understand the underlying physics of the particle settling modification by the turbulent carrier phase.

Rosa et al. \cite{rosa2016settling} also found a linear hindering behavior, consistent with our findings of section \ref{sc:ganal} , with a slope close to -0.3, when the lateral movement of the particles was suppressed in direct numerical simulations (figure \ref{fig-app-3}).

\label{sc:scalings2}

\bibliography{SETTLING-APS.bib}
\end{document}